\documentclass[namedreferences]{SolarPhysics}
\pdfoutput=1
\usepackage[%
optionalrh,%
natbib,%
]{spr-sola-addons} %

\usepackage{solaheader}	%
\newcommand{\solareceiveddate}{2 September 2011} 
\newcommand{\solaaccepteddate}{12 March 2012}
\newcommand{\solaonlinedate}{5 April 2012}
\newcommand{\Xdoi}{s11207-012-9974-z}  
\renewcommand{\solavolume}{X}
\renewcommand{\firstpage}{1}
\renewcommand{\lastpage}{26}
\renewcommand{\solayear}{2012}
\usepackage[pdftex]{graphicx}            %
\usepackage{color}                       %
\usepackage{url}                         %

\newcommand{\etal}{{\it et al.}}

\newcommand{\ion}[2]{\mbox{{#1}~\small\rmfamily\scshape{#2}}}
\newcommand{\kms}{km~s$^{-1}$}
\newcommand\arcsec{\mbox{$^{\prime\prime}$}}%

\newcommand{\aap}{    {\it Astron. Astrophys.}}
\newcommand{\aaps}{   {\it Astron. Astrophys. Suppl.}}

\newcommand{\apj}{    {\it Astrophys. J.}}
\newcommand{\apjs}{   {\it Astrophys. J. Supp.}}
\newcommand{\apjl}{   {\it Astrophys. J. Lett.}}

\newcommand{\grl}{    {\it Geophys. Res. Lett.}}

\newcommand{\jgr}{    {\it J. Geophys. Res.}}

\newcommand{\solphys}{{\it Solar Phys.}}

\newcommand{\ssr}{    {\it Space Sci. Rev.}}
\newcommand{\bain}{   {\it Bull. Astr. Inst. Netherlands}}
\newcommand{\araa}{   {\it Ann. Rev. Astron. Astrophys.}}
\begin{document}

\begin{article}

\begin{opening}

\title{%
  Coronal Diagnostics from Narrowband Images around 30.4 nm
}

\author{%
  V.~\surname{Andretta}$^{1}$\sep
  D.~\surname{Telloni}$^{2}$\sep
  G.~\surname{Del Zanna}$^{3}$      
}

\runningauthor{Andretta \etal}
\runningtitle{Coronal Diagnostics from 30.4 nm Images}

\institute{%
  $^{1}$ INAF/Osservatorio Astronomico di Capodimonte, 
  Salita Moiariello 16, 80131 Napoli, 
  Italy --
  email: \url{andretta@oacn.inaf.it}\\ 
  $^{2}$ INAF/Osservatorio Astrofisico di Torino, 
  Strada Osservatorio 20, 10025 Pino Torinese (TO),
  Italy --
  email: \url{telloni@oato.inaf.it} \\
  $^{3}$ DAMTP, Centre for Mathematical Sciences, University of Cambridge, UK
  email: \url{G.Del-Zanna@damtp.cam.ac.uk}
}

\begin{abstract}
  Images taken in the band centered at 30.4~nm are routinely used to map the
  radiance of the \ion{He}{ii} Ly\,$\alpha$ line on the solar disk.  That line
  is one of the strongest, if not the strongest, line in the EUV
  observed in the solar spectrum, and one of the few lines in that wavelength
  range providing information on the upper chromosphere or lower transition
  region.  However, when observing the off-limb corona the contribution from
  the nearby \ion{Si}{xi} 30.3~nm line can become significant.  In this work
  we aim at estimating the relative contribution of those two lines in the
  solar corona around the minimum of solar activity.  We combine measurements
  from CDS taken in August 2008 with temperature and density profiles from
  semiempirical models of the corona to compute the radiances of the two
  lines, and of other representative coronal lines (\textit{e.g.}, \ion{Mg}{x} 62.5~nm,
  \ion{Si}{xii} 52.1~nm).  Considering both diagnosed quantities from line
  ratios (temperatures and densities) and line radiances in absolute units, we
  obtain a good overall match between observations and models.  We find
  that the \ion{Si}{xi} line dominates the \ion{He}{ii} line from just
  above the limb up to $\approx 2\, R_\odot$ in
    streamers, while its contribution to
  narrowband imaging in the 30.4~nm band is expected to become smaller, even
  negligible in the corona beyond $\approx 2$ -- $3\, R_\odot$, the
    precise value being strongly dependent on the coronal temperature profile.
\end{abstract}


\end{opening}
\section{Introduction}\label{sec:introduction}

Narrowband imaging of the Sun in the EUV has become an almost indispensable
tool for solar physicists.  Instruments like the \textit{Extreme UltraViolet
  Imaging Telescope} (EIT) onboard the \textit{Solar and Heliospheric
  Observatory} (SOHO), and the \textit{Atmospheric Imaging Assembly} (AIA) onboard 
the \textit{Solar Dynamics Observatory} (SDO) routinely monitor the Sun at various
wavelengths.  Both instruments observe, among others, in bands centered around
30.4~nm, the wavelength of the Ly\,$\alpha$ line of ionized helium.

Observations with the \textit{Sounding CORona Experiment} \citep[SCORE: ][]%
{Fineschi-etal:04,Fineschi:06}, flown on HERSCHEL, a NASA sounding-rocket
payload, also include such a band.  Finally, 
the \textit{Multi Element Telescope for Imaging and Spectroscopy} (METIS),
one of the scientific payloads selected for \textit{Solar Orbiter}, the
M-class mission of the ESA Cosmic Vision program for the exploration of the Sun at
a distance as close as 0.28 AU, will also have a spectro-imaging
channel at that wavelength.

\subsection{The Spectral Region around 30.4 nm}

The most interesting line in the band around 30.4 nm is certainly the
\ion{He}{ii} 30.4~nm, the Ly\,$\alpha$ of ionized helium.  Narrowband imaging
of the solar disk around that line therefore allows probing relatively cool
plasma ($T < 10^5$~K) in the EUV, although the details of the formation of
that line in the quiescent solar atmosphere are still somewhat controversial
\citep[\textit{e.g.},][]{Andretta-etal:03,Pietarila-Judge:04}.

However, for the extended corona, one of the most interesting
diagnostics possible with the \ion{He}{ii} 30.4~nm line is related to a
radiative ``pumping'' mechanism analogous to that exhibited by hydrogen
Ly\,$\alpha$.  Furthermore, the nearby
strong \ion{Si}{xi} resonance line  at 30.3~nm %
adds the possibility of probing radial velocities in the range of $\approx
400$~\kms\ via a similar mechanism.

\subsection{The Ratio \ion{Si}{xi} 30.3~nm/\ion{He}{ii} 30.4~nm}

\begin{figure}[b!th]
  \centering
  \includegraphics[trim=50 100 50 60,clip=false,width=0.95\linewidth]%
  {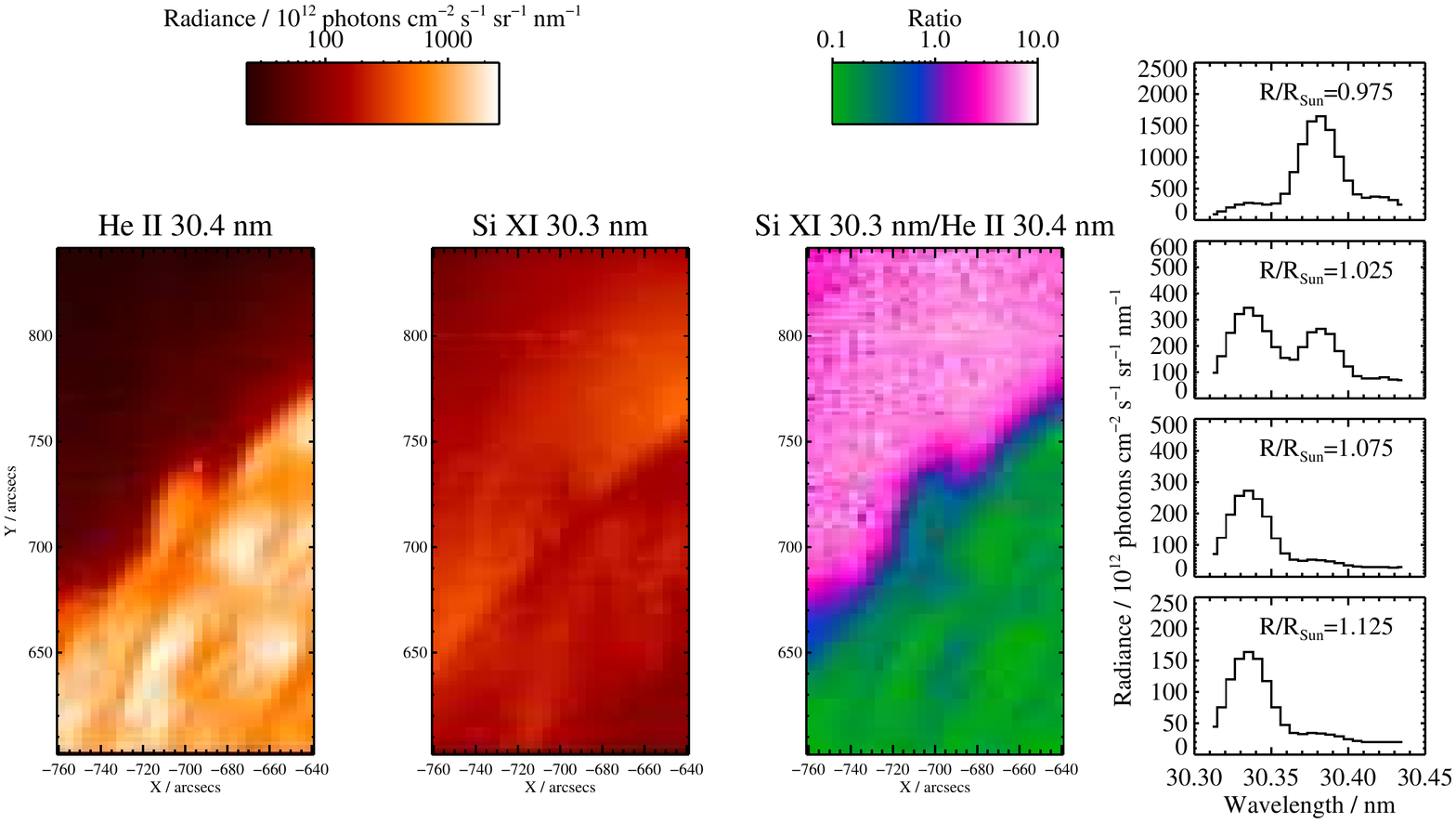}
  \caption{%
    Images (leftmost two panels) of the \protect\ion{He}{ii} 30.4~nm and
    \protect\ion{Si}{xi} 30.3~nm lines created from CDS spectral rasters, and
    their ratio.  In the third panel, ratios coded with red/white indicate a
    high value of \protect\ion{Si}{xi} 30.3~nm/\protect\ion{He}{ii} 30.4~nm,
    while green indicate a dominant \protect\ion{He}{ii} 30.4~nm line.  The
    average spectra around 30.4~nm at different heliocentric distances from
    the same data set are shown in the rightmost panels.
  }
  \label{fig:nis_sample}
\end{figure}

The presence of the nearby \ion{Si}{xi} line, while providing 
additional diagnostic opportunities, 
poses a problem for the interpretation of
narrowband images.  The ratio \ion{Si}{xi} 30.3~nm/\ion{He}{ii} 30.4~nm is
around 0.1 in the quiescent solar atmosphere seen on the solar disk
\citep{DelZanna-Andretta:11}, or even smaller in active regions
\citep{Thomas-Neupert:94}, but dramatically changes in the corona above the
limb.

To illustrate this, we show in Figure \ref{fig:nis_sample} the peak radiances
of the two lines, measured from spectra taken with the \textit{Normal
  Incidence Spectrograph} (NIS) of the SOHO/\textit{Coronal Diagnostic
  Spectrometer} \citep[CDS:][]{Harrison-etal:95}.  The raster scans shown here were
taken on 21 January 1998, and can be considered representative of relatively
quiescent off-limb regions: only a small prominence is visible in the
field of view (FOV).  The ratio of the two lines is also shown.

The rapid increase of the ratio \ion{Si}{xi} 30.3~nm/\ion{He}{ii} 30.4~nm
above the limb is also apparent from the average profiles at various heights
above the limb shown in Figure \ref{fig:nis_sample}.  However, data obtained
from CDS/NIS are limited to the lower corona; it is entirely possible that
further out in the corona the ratio \ion{Si}{xi} 30.3~nm/\ion{He}{ii} 30.4~nm
could decrease again.  Whether and where that happens is of utmost
relevance for the interpretation of narrowband imaging of the outer corona.

In this article we will make an estimate of the ratio of the two lines in
typical conditions of the corona at solar minimum.  Using observed EUV,
spectrally resolved radiances from SOHO/CDS (Section \ref{sec:usun}), we will
first estimate average on-disk and off-limb radiances in several coronal
lines, including those for \ion{Si}{xi} 30.3~nm and \ion{He}{ii} 30.4~nm; from the line
ratios we will also derive average densities and temperatures in the corona.
These measurements will then provide the input disk irradiance needed 
to calculate the radiative excitation of the coronal 
lines, as well as constraints to the calculations at the lower boundaries of
the models, \textit{i.e.} near the solar limb (Section \ref{sec:calcs}).

\section{SOHO/CDS Radiances}\label{sec:usun}

In order to derive both mean on-disk and off-limb radiances of lines in the
spectral range observed by CDS/NIS, we used whole-Sun scans taken with the
CDS/NIS study called USUN in the CDS database.  

The CDS USUN study consists of 69 rasters, for a total of 700 to 1000
exposures with the 4\arcsec\ slit, covering the whole Sun and part of the
off-limb corona in about 13 hours.  Further details on the CDS USUN study 
are given in \cite{Thompson-Brekke:00} and \cite{DelZanna-Andretta:11}.  
Here we use the data processed and analyzed as described in detail in
the latter article, including the radiometric calibration of CDS/NIS of
\cite{DelZanna-etal:01,DelZanna-etal:10}.
We note that the new calibration is significantly
different from the previous ones.  We also note that
measurements of the \ion{He}{ii} 30.4~nm radiance/irradiances 
have been largely inconsistent throughout the literature, and that 
the \cite{DelZanna-Andretta:11} calibration provides irradiances
in agreement with the SDO \textit{Extreme Ultraviolet Variability Experiment}
(EVE) prototype measurements of 14 April 2008 as described in the same article.
From comparisons between our measurements with those
by various instruments, 
two  errors in the analysis software of  CDS and the 
\textit{EUV Normal-Incidence  Spectrometer} (EUNIS: \citealt{jordan_brosius:07})
were uncovered \citep{wang-etal:11}. Our CDS measurements are now also in 
agreement with the EUNIS ones, within a few percent.

\begin{figure}[htb]
  \centering
  \includegraphics[trim=40 150 40 170,clip=false,width=0.95\linewidth]%
  {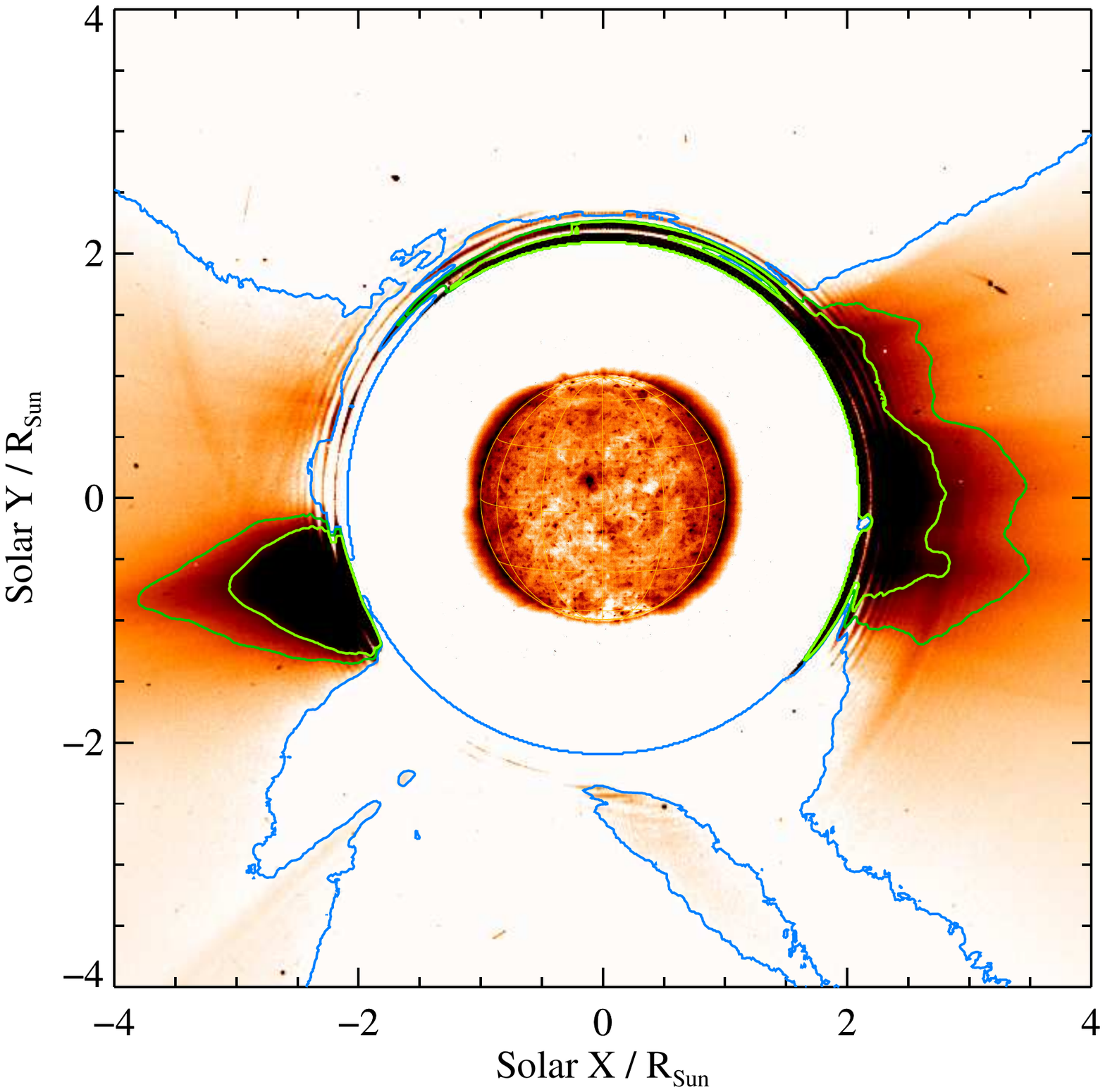}
  \caption{%
    Image of the solar corona from SOHO/LASCO-C2, taken on 
    25 August 2008 07:29:36 UT,
    overlaid on a SOHO/EIT 195 image taken on the same day at 07:13:47 UT. A
    radial average of the corona has been subtracted from the LASCO image to
    enhance the contrast of coronal structures.
  }
  \label{fig:lasco_eit}
\end{figure}

The  CDS USUN 
  scans have been taken approximately once a month starting in 1998. We
chose to analyze a USUN scan taken on 25 August 2008, \textit{i.e.}
during the extended minimum between activity Cycles 23 and 24.  The
configuration of the corona from SOHO/EIT
\citep{Delaboudiniere-etal:95} and from the C2 telescope of the \textit{Large Angle and Spectrometric Coronagraph}
\citep[LASCO:][]{Brueckner-etal:95} on that date is shown in
Figure \ref{fig:lasco_eit}.  For display clarity, an average radial profile has
been subtracted from the LASCO-C2 image.

\subsection{Selected Regions of the FOV}\label{sec:usun:roi}

\begin{figure}[htb]
  \centering
  \includegraphics[trim=60 100 40 100,clip=false,width=0.95\linewidth]%
  {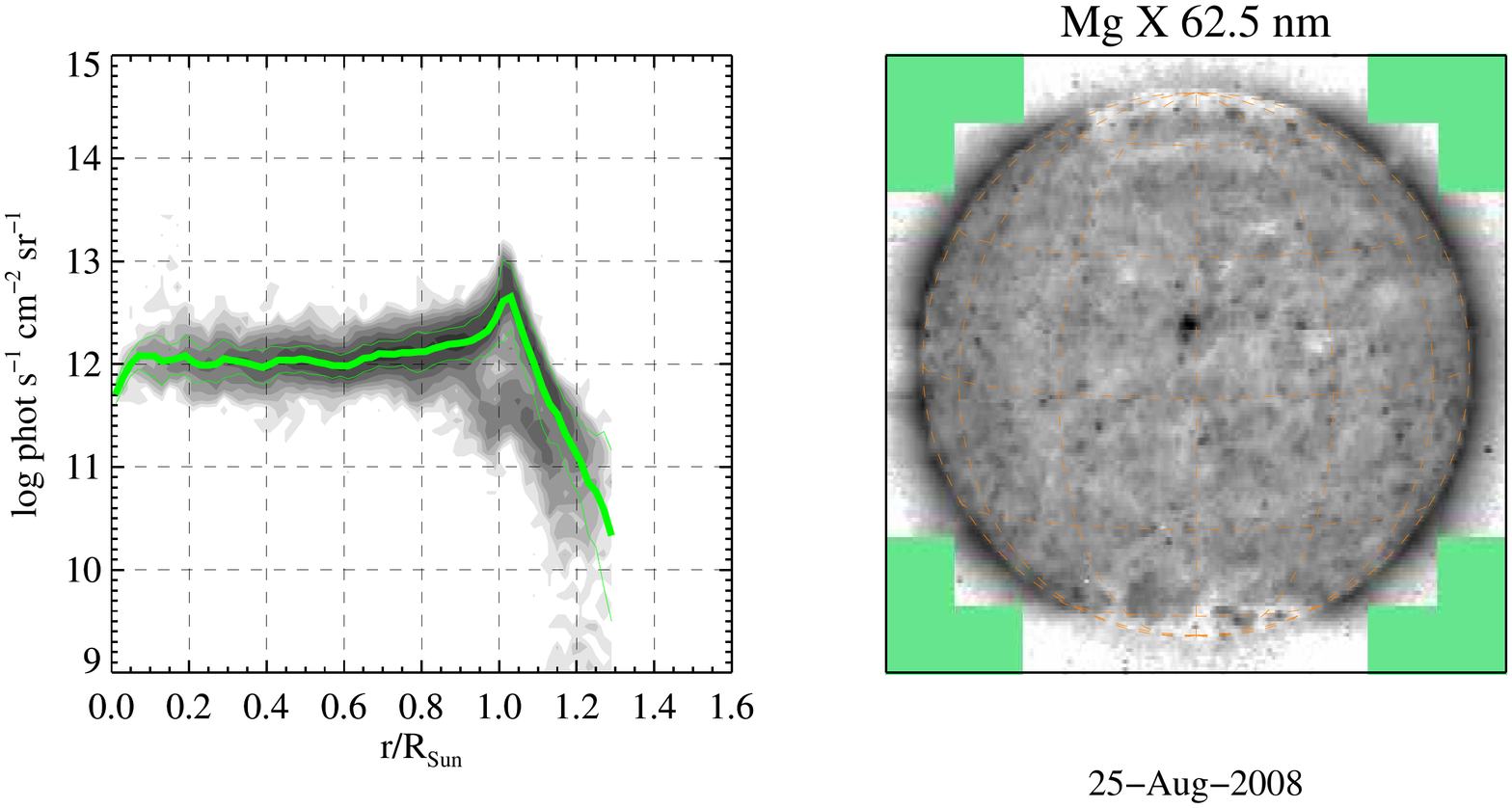}%
  \caption{%
    Right: Radiance of the \protect\ion{Mg}{x} 62.5~nm line from
    CDS/NIS observations; USUN study on 25 August 2008. Left: Histogram
    showing the distribution of radiances in the line as a function of
    heliocentric distance; green: median value (plus or minus one standard
    deviation).
  }
  \label{fig:cds_usun}
\end{figure}

For each line in the CDS spectral range, it is possible to build a
monochromatic image of the Sun.  An example is shown in the right-hand panel
of Figure \ref{fig:cds_usun}, for the \ion{Mg}{x} 62.5~nm line.  The solar disk
on that date did not exhibit significant activity, while two symmetrical polar
coronal holes are clearly visible.

For each monochromatic image, we then built a histogram of radiances as
function of distance from the center of the Sun.  An example of the
two-dimen\-sional histogram thus obtained is shown in the left-hand panel of
Figure \ref{fig:cds_usun}.  We then obtained an average radiance radial profile
from the median of the histograms at each heliocentric distance (thicker line
in Figure \ref{fig:cds_usun}).  The standard deviation of the profile is also
shown (thinner lines).

\begin{figure}[hbt]
  \centering
  \includegraphics[trim=60 100 40 100,clip=false,width=0.95\linewidth]%
  {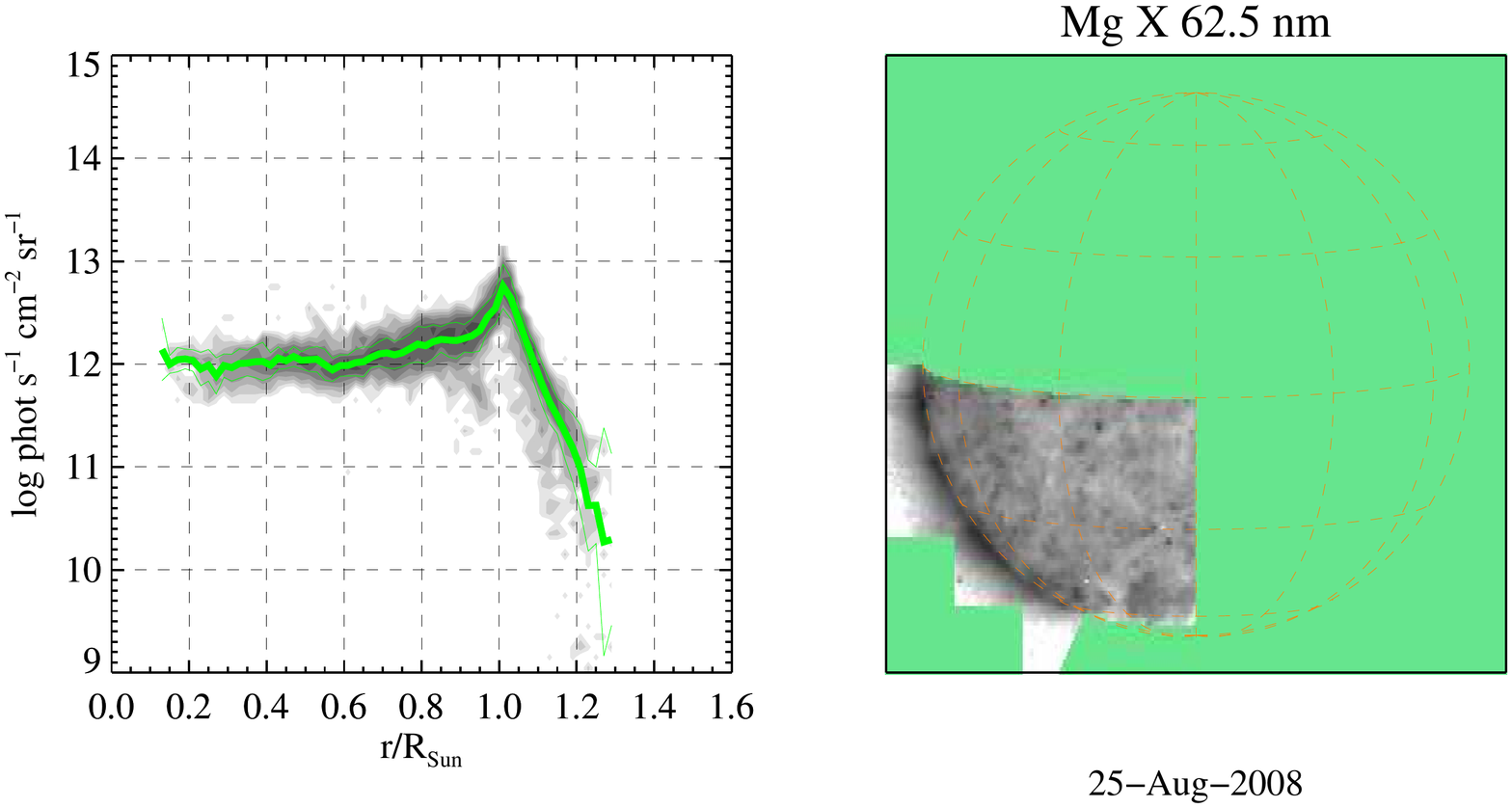}
  \includegraphics[trim=60 100 40 100,clip=false,width=0.95\linewidth]%
  {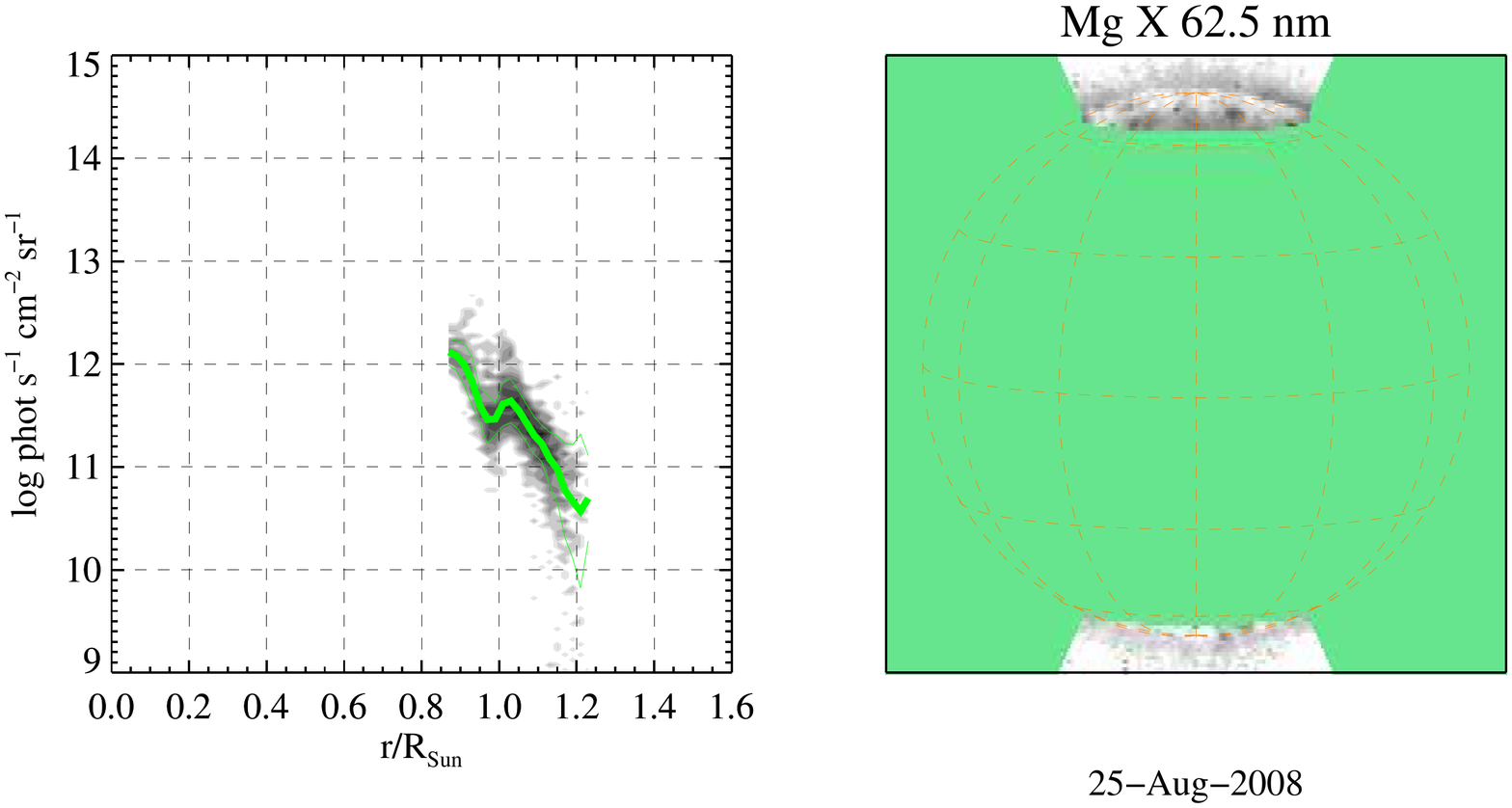}
  \caption{%
    Right panels: Radiance of the \protect\ion{Mg}{x} 62.5~nm line from
    CDS/NIS observations, for selected regions of the FOV; USUN study on 25
    August 2008. 
    Left panels: Histograms
    showing the distribution of radiances in the line as a function of
    heliocentric distance; green: median value (plus or minus one standard
    deviation).
    Top panels: southeast quadrant,
    roughly corresponding to the base of the streamer shown in 
    Figure \ref{fig:lasco_eit};
    Bottom panels: polar regions (coronal holes).
  }
  \label{fig:cds_usun_regions}
\end{figure}

The same procedure can be applied to specific regions of the FOV.  In
particular, we selected the southeast quadrant of the disk, corresponding to the base
of the streamer visible in Figure \ref{fig:lasco_eit}, and the two polar
regions.  The corresponding images and radiance histograms are shown in
Figure \ref{fig:cds_usun_regions}.  We have verified that the choice of a
specific region of interest outside the polar coronal holes does not affect
significantly the radiance radial profiles for most lines, a fact consistent
with the very low activity state of the lower corona on that date.

\subsection{Off-limb Radiances}\label{sec:usun:offlimb}

There
are at least two known sources of spurious radiation that can affect the
measurement of off-limb line radiances in CDS.  

First, radiation from the much brighter solar disk can be diffused into the
tail of the instrument's point-spread function (PSF) to contribute to off-limb
line radiation.  \cite{Harrison-etal:95} give values for the PSF (measured at
6.8~nm) of $10^{-6.2}$ and $10^{-6.6}$ for distances of 100\arcsec\ and
150\arcsec, respectively, from the core.  Based on those figures, at distances
of $r\approx 1.2\; R_\odot$ we would expect the fractional contribution of the
disk to line radiances to be well below $10^{-6}$, compared to a decrease of
the order of little more than a factor ten in typical coronal lines 
(Figures \ref{fig:cds_usun} and \ref{fig:cds_usun_regions}).  However, the
in-flight performance of the instrument is not necessarily as good as
indicated by those laboratory measurements, especially after the loss of
contact with the SOHO spacecraft in June 1998, which led to a significant
alteration of the instrumental response.

\begin{figure}[hbt]
  \centering
  \includegraphics[trim=50 35 50 35,clip=false,width=0.95\linewidth]%
  {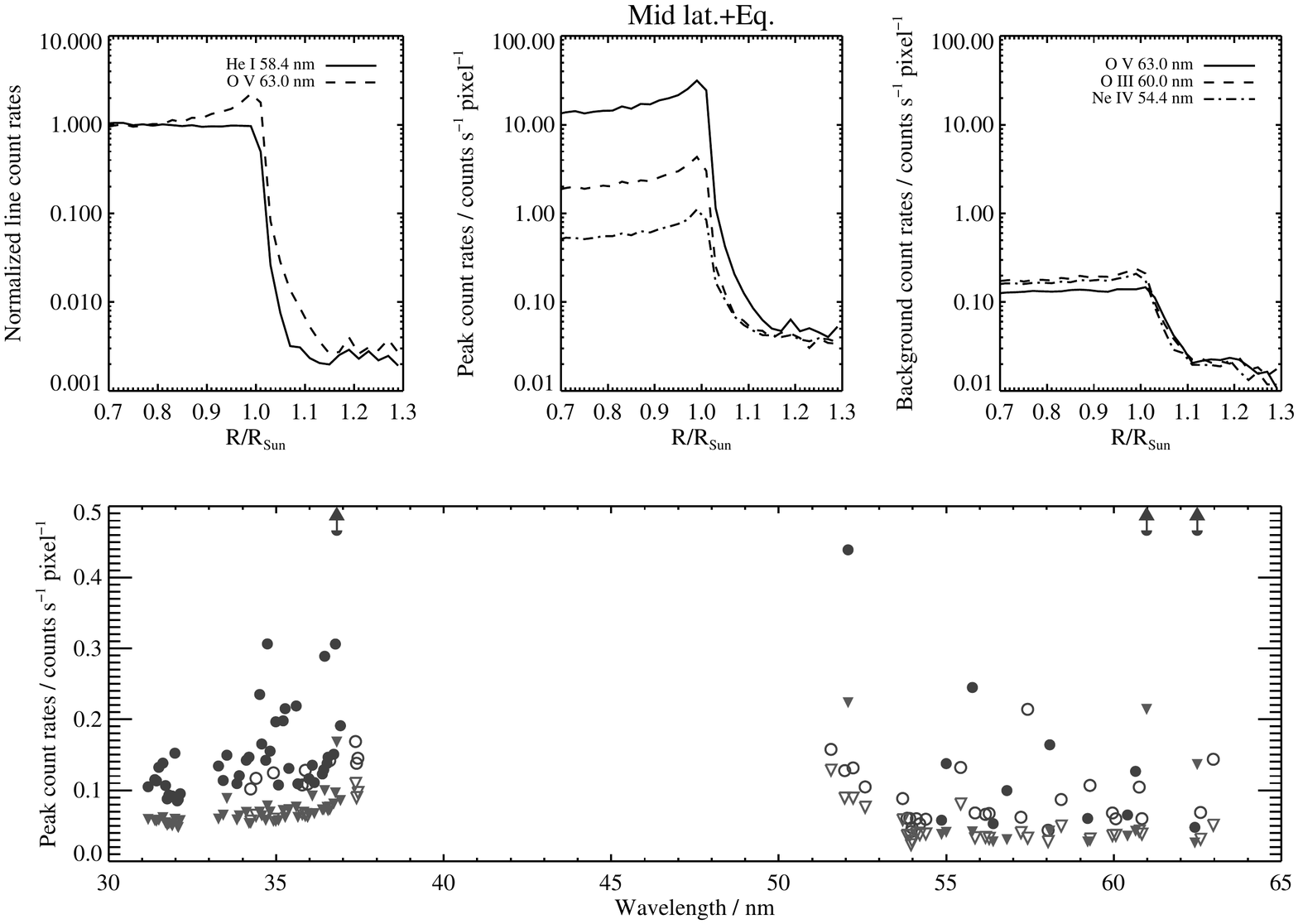}
  \caption{%
      Radial profiles for the count rates of two of the strongest lines in
      the CDS range, averaged over equatorial and mid-latitude regions, and
      normalized to the mean line count rate over the solar disk
      (leftmost, upper panel). The other two plots in the upper row
      show, for three representative lines in the NIS 2 range with formation
      temperature below $5\times 10^5$~K, the average count rates in the line
      center pixel (center, upper panel), and in adjacent (background)
      pixels (rightmost, upper panel).  The bottom panel shows the peak
      count rate per pixel for all of the lines detected in the CDS range \textit{vs.}\
      wavelength, averaged in the range $1.05<r/R_\odot<1.15$ (circles) and
      $r/R_\odot>1.2$ (triangles).  Open and filled symbols mark lines with
      formation temperature below and above $5\times 10^5$~K, respectively.
      The values for the strongest coronal lines (the \ion{Mg}{x} doublet at
      60.9 and 62.5~nm, \ion{Mg}{ix} 36.8~nm) around $r/R_\odot=1.1$ are off
      the vertical scale, and thus are not shown here.
  }
  \label{fig:cds_straylight}
\end{figure}

To estimate the contribution of disk scattered radiation, we examined the
off-limb behavior of the strongest cool lines (lines with formation
temperature much less than $10^6$~K) in the CDS range: \ion{He}{i}~58.4~nm and
\ion{O}{v}~62.9~nm, assuming that the coronal contribution to their off-limb
measured radiances is negligible (Figure \ref{fig:cds_straylight}, left panel in
the upper row).  From our data, it appears that the fractional contribution of
disk scattered radiation is below $3\times 10^{-3}$ at $r\approx 1.2\; R_\odot$.

A more precise determination of this contribution is impeded by the presence
of a continuum background, only weakly dependent on wavelength
\citep{Thompson-Brekke:00}.  
The source of
this background is unclear; however, if of solar origin, its pattern of spatial
variation on the disk would suggest a cool source
(Figure \ref{fig:cds_straylight}, center and right panels in the upper row).
According to \cite{Thompson-Brekke:00}, for instance, such a background is
mainly due to Ly\,$\alpha$ radiation scattered by the far wings of the grating
profile, but with likely contributions from a variety of other lines.  
In
any case, the presence of that background can hamper accurate measurement of
off-limb lines, especially of the faintest ones.

The average line peak count rates at $r\approx 1.25\; R_\odot$ (average above
$1.2\; R_\odot$), shown in the lower panel of Figure \ref{fig:cds_straylight}
(triangles), clearly outline the background continuum; only a few strong lines
are still measurably above that level.  In contrast, at $r\approx 1.1\;
R_\odot$ (average between $1.05$ and $1.15\; R_\odot$), most lines with
formation temperature above $\approx 10^6$~K are still clearly detectable above
the background (filled circles in Figure \ref{fig:cds_straylight}).  For 
this discussion, we are only interested in a few, relatively
strong lines (see following sections); those lines are still measurable
up to at least $r\approx 1.2\; R_\odot$ or beyond (as is the case for the
\ion{Mg}{x} 62.5~nm, \ion{Mg}{ix} 36.8~nm, or \ion{Si}{xii} 52.1~nm lines).  The
same analysis can be applied to polar coronal holes, leading to similar
results, although with larger uncertainties due to lower count rates in
coronal lines.

\begin{figure}[bth]
  \centering
  \includegraphics[angle=180,trim=50 35 100 60,clip=false,width=0.95\linewidth]%
  {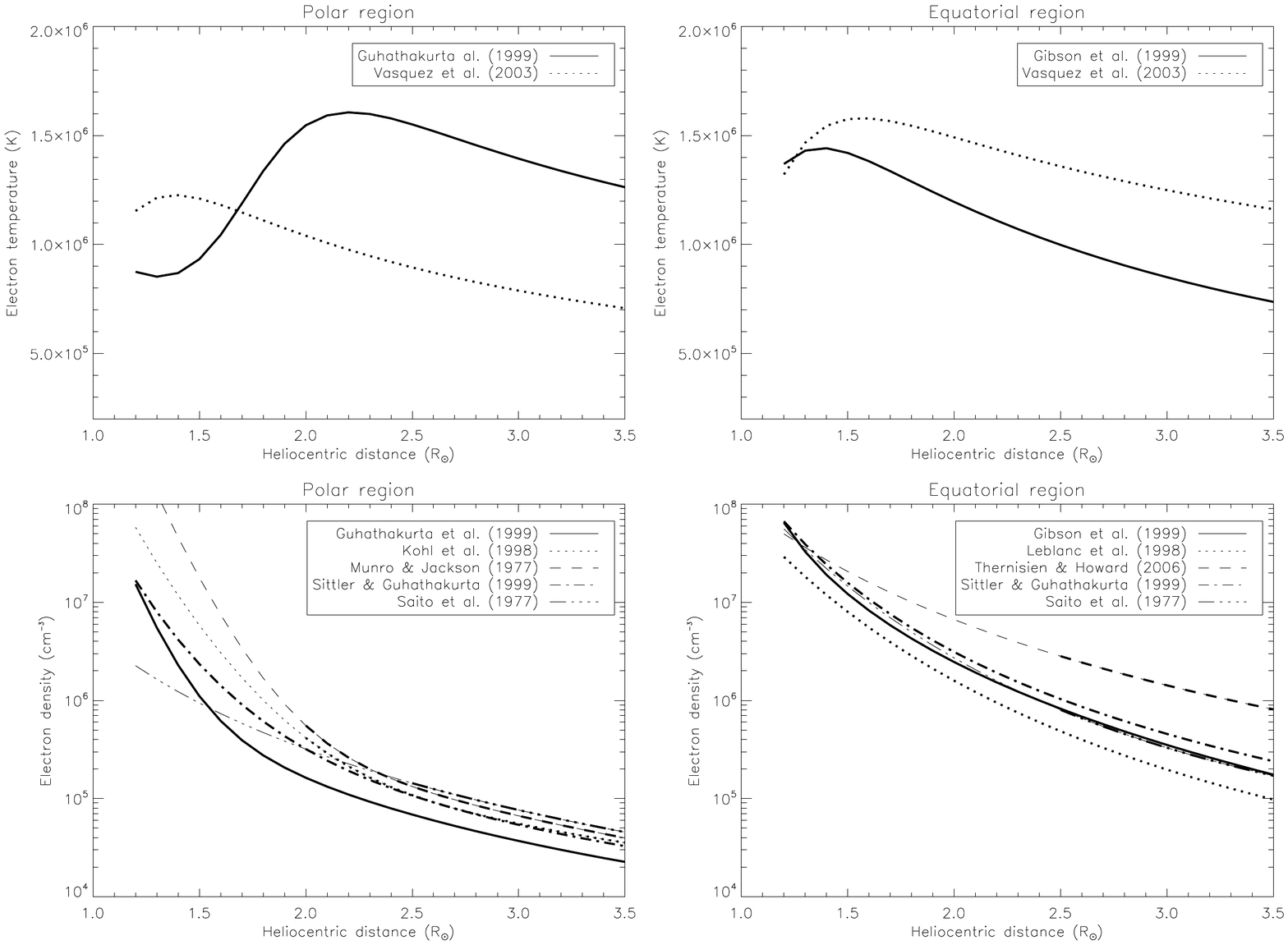}%
  \caption{%
    Electron temperature (top panels) and density (bottom
      panels) models in polar coronal holes (left panels) and in
    equatorial streamers (right panels); the lines representing the
    different models are thicker in the corresponding domain of validity.%
  }
  \label{fig:electron_temperature_density}
\end{figure}

As mentioned in Section \ref{sec:introduction}, these measured radiances will be
used to constrain various semiempirical models of the corona found in
literature.  A summary of the plasma parameters from those models is shown in
Figure \ref{fig:electron_temperature_density}; the models are described
in more detail in Section \ref{sec:calcs}.

\subsection{Temperatures and Densities from CDS
  Radiances}\label{sec:usun:diagnostics}

From the average radiance radial profiles, it is  possible to obtain
estimates of average temperatures and densities above the limb 
 using line-ratio diagnostics.  We selected the density-sensitive ratio
\ion{Si}{ix} 34.2/35.0~nm 
  and the temperature-sensitive ratio \ion{Mg}{x} 62.5~nm/\ion{Mg}{ix}
  36.8~nm, in analogy with the similar analyses of \cite{Gibson-etal:99} and
  \cite{fludra_etal:99}, based on SOHO/CDS measurements during the solar minimum
  in 1996.
The latter temperature-sensitive ratio has the advantage that the lines are bright and from the same
element. 

The densities and temperatures from the observed line ratios were obtained
using the CHIANTI database \citep{Dere-etal:97}, version 6
\citep{Dere-etal:09}, in particular including the new ion abundances
in ionization equilibrium.

\begin{figure}[htb]
  \centering
  \includegraphics[trim=60 50 40 35,clip=false,width=0.475\linewidth]
  {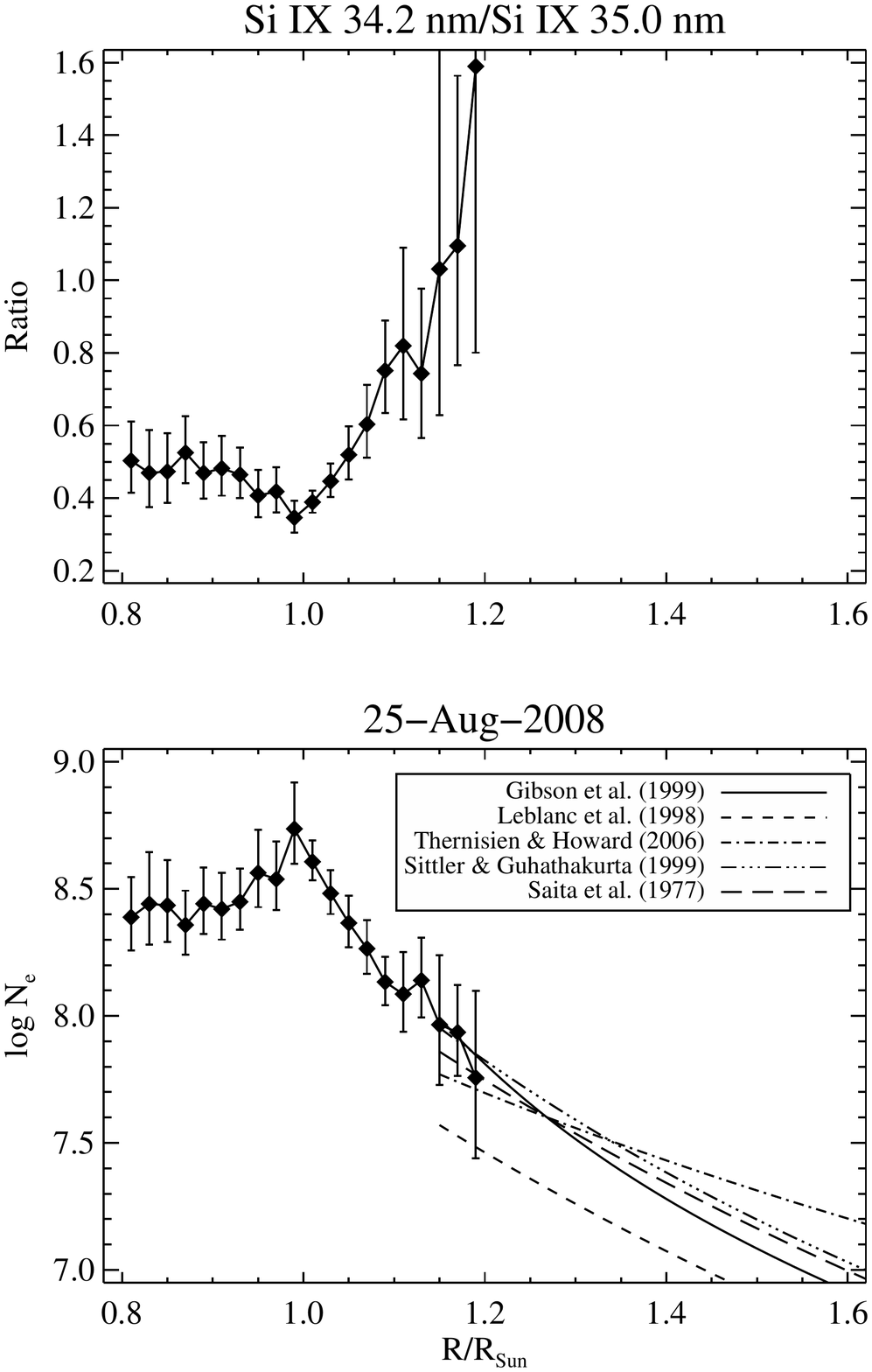}
  \includegraphics[trim=60 50 40 35,clip=false,width=0.475\linewidth]
  {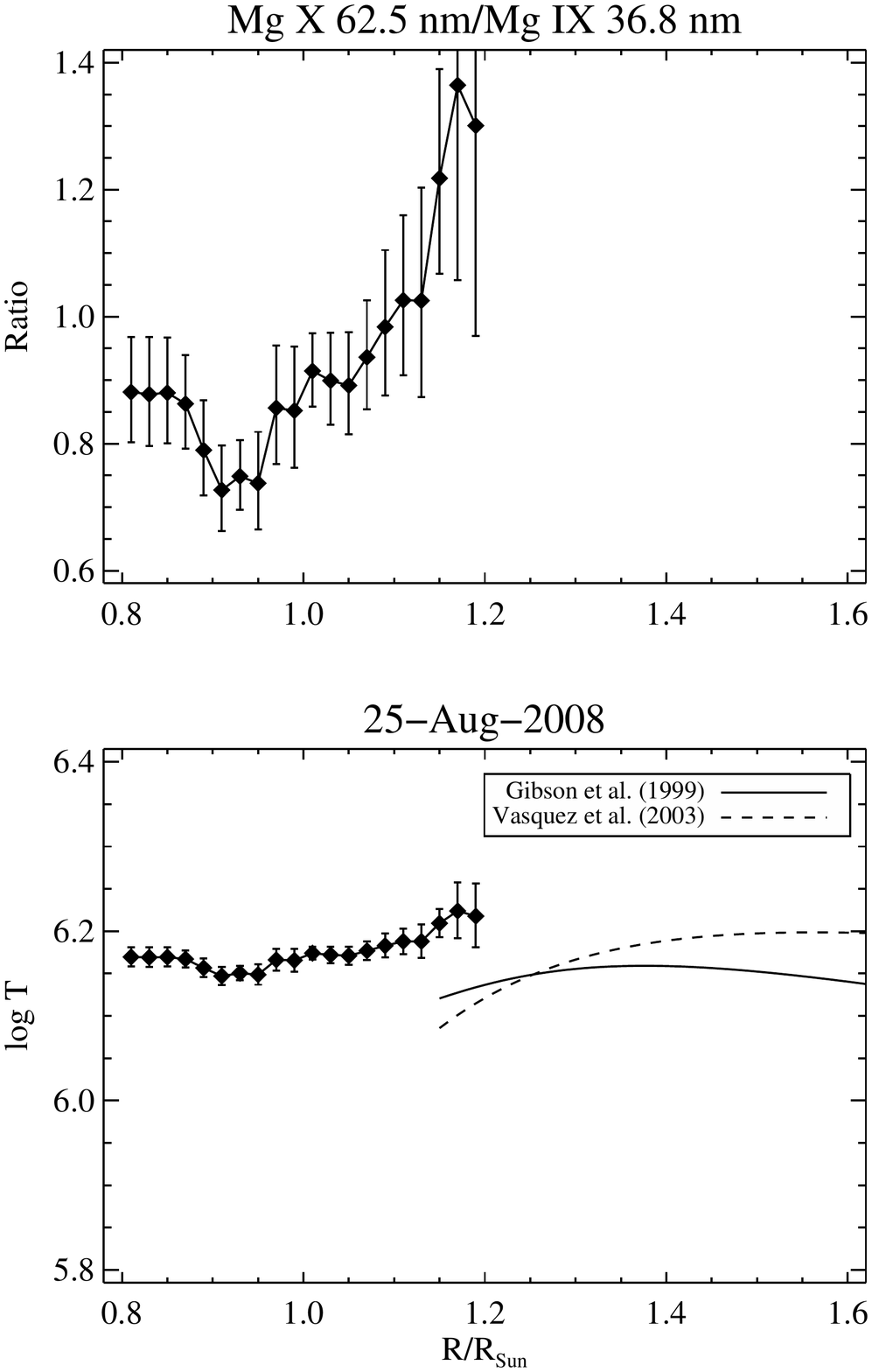}%
  \caption{%
    Top panels: Mean line ratios from CDS observations, for the southeast
    region shown in Figure \ref{fig:cds_usun_regions}. %
    Bottom panels: Corresponding diagnosed quantities (left
      panels: $N_\mathrm{e}$, right panels: $T_\mathrm{e}$), compared
    with various semiempirical models of streamer regions.
  }
  \label{fig:cds_ratios_eq}
\end{figure}
\begin{figure}[hbt]
  \centering
  \includegraphics[trim=60 50 40 35,clip=false,width=0.475\linewidth]
  {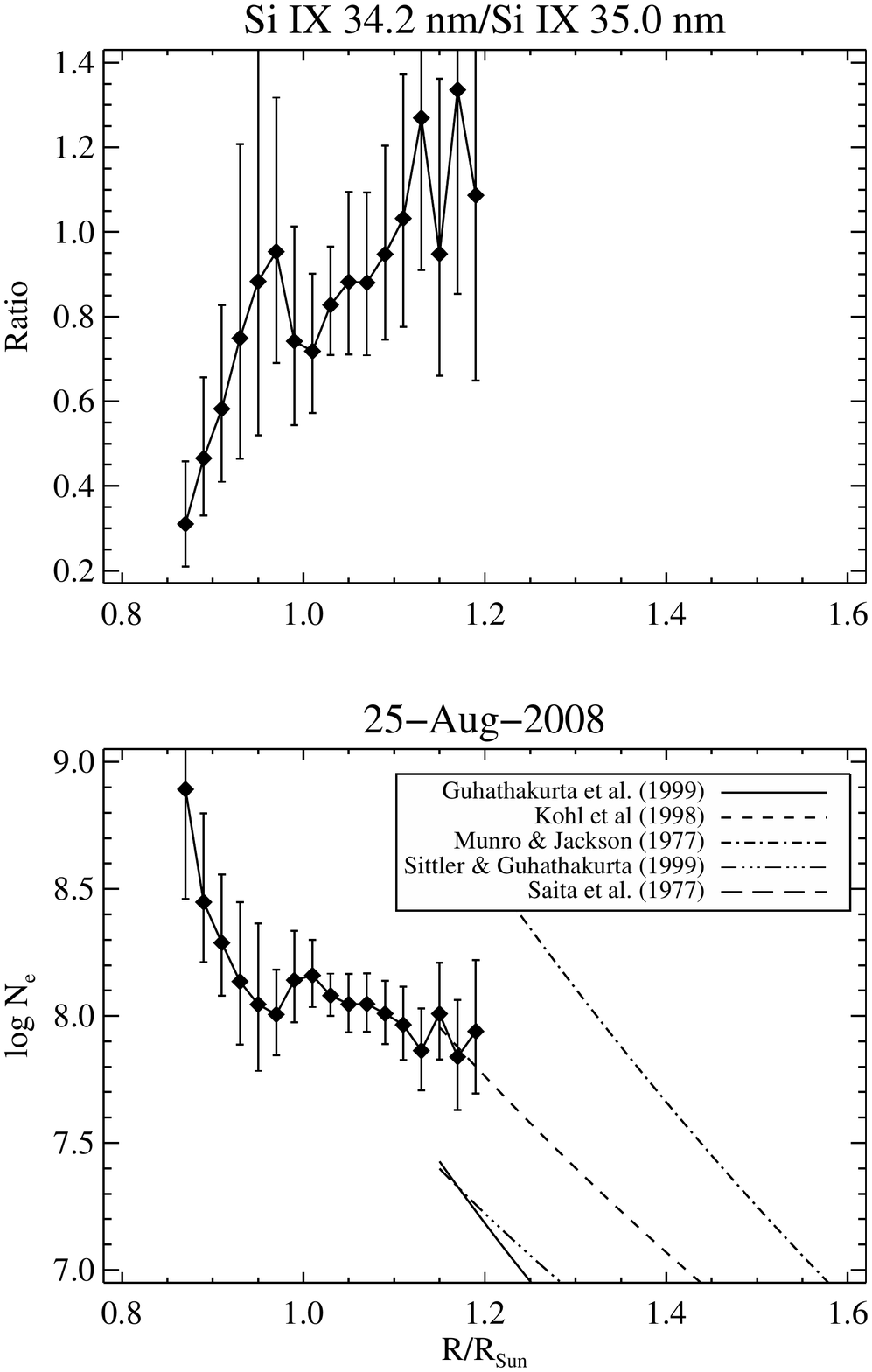}%
  \includegraphics[trim=60 50 40 35,clip=false,width=0.475\linewidth]
  {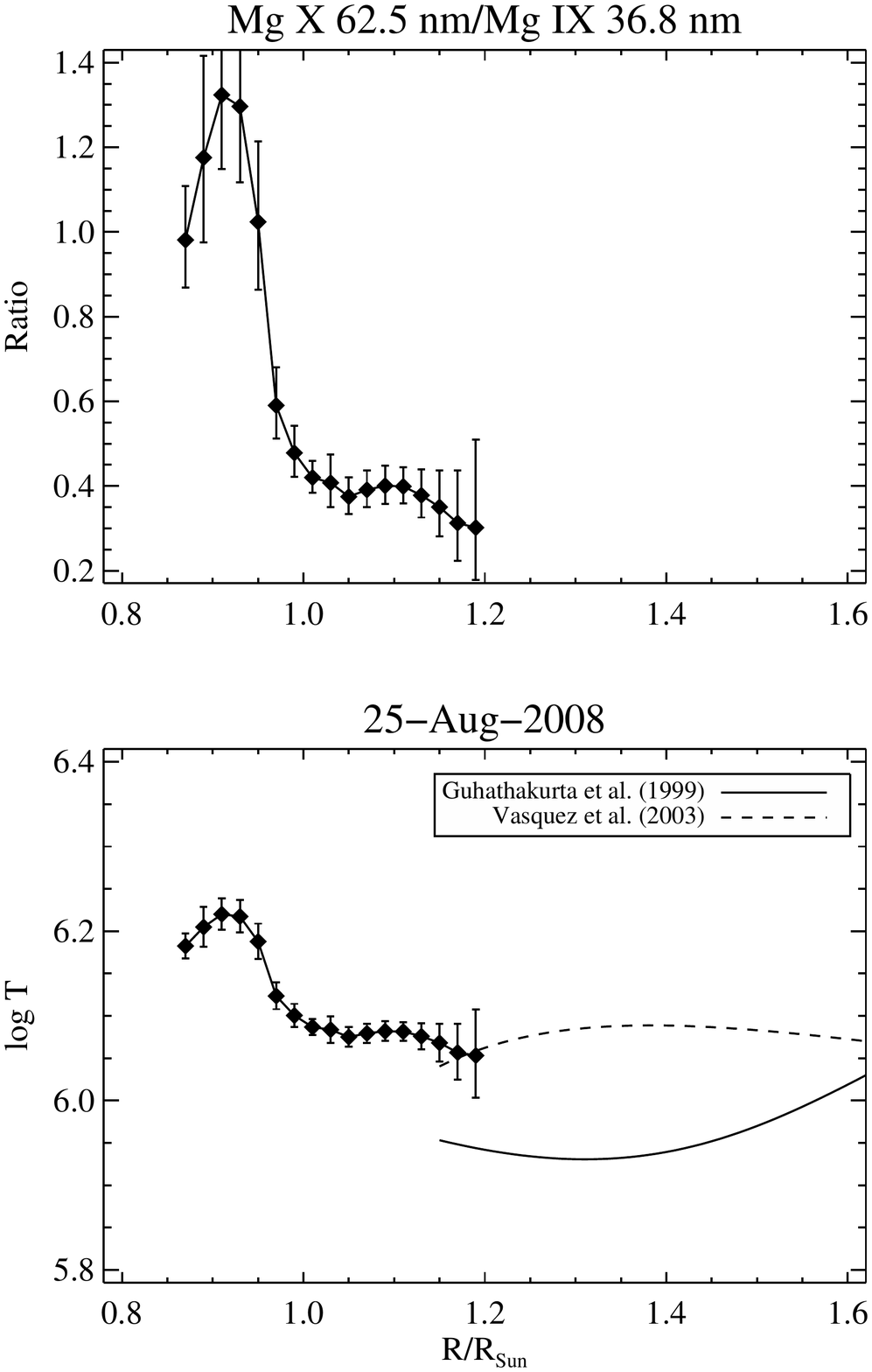}%
  \caption{%
    Top panels: Mean line ratios from CDS observations, for the the polar
    regions shown in Figure \ref{fig:cds_usun_regions}. %
    Bottom panels: Corresponding diagnosed quantities (left
      panels: $N_\mathrm{e}$, right panels: $T_\mathrm{e}$), compared
    with various semiempirical models of polar (coronal hole) regions.
  }
  \label{fig:cds_ratios_ch}
\end{figure}

  The line ratios measured at the base of the southeast streamer are shown in
  Figure \ref{fig:cds_ratios_eq} (top panels), while the corresponding plasma
  parameters, compared with density and temperature profiles from various
  authors, are shown in the lower panels. Figure \ref{fig:cds_ratios_ch} reports
  the same results for the polar coronal holes. 

  The work of \cite{Gibson-etal:99} and \cite{fludra_etal:99} actually base
  their temperature measurement on the ratio of the lines \ion{Si}{xii}
  52.1~nm and \ion{Mg}{x} 62.5~nm.  However, those two
  lines may be affected by an anomalous behavior of the Li-like ions
  in many emission measure analyses, as described in
  \cite{delzanna_etal:02_aumic}.  Indeed, the ratio measured in our data is at
  variance both with the other ratio and with the literature profiles. In
  polar coronal holes, in particular, that ratio yields too high a
  temperature, even slightly higher than in the rest of the corona.  
  We used a photospheric mixture of abundances, as discussed in
  Section \ref{sec:calcs:corona:abundances}, but since both Mg and Si are
  elements with low first ionization potential (\,FIP\,), adopting ``coronal''
  abundances would not significantly reduce such a discrepancy.
  This incongruity may be due
  to the anomalous behavior of Li-like ions mentioned above and discussed in
  Section \ref{sec:comparison:eq}, but also to the fact that the \ion{Si}{xii}
  52.1~nm line is weaker than the \ion{Mg}{x} 62.5~nm line, much more so in coronal
  holes, and thus may possibly be affected by the stray light described in
  Section \ref{sec:usun:offlimb}.  With these caveats in mind, while regarding
  the temperature from the ratio of the two as unreliable, we will
  nevertheless consider for both lines the radiance radial profiles and
  compare them with the corresponding calculations of
  Section \ref{sec:comparison}.

\subsubsection{Streamer}

The density profile for the off-limb quiescent corona from the ratio
\ion{Si}{ix} 34.2/ 35.0~nm, if extrapolated to heliocentric distances $r>1.2\;
R_\odot$, is in excellent agreement with the values listed by
\cite{Gibson-etal:99}, as well as with the density profiles of
\cite{Thernisien-Howard:06}, \cite{Sittler-Guhathakurta:99}, and
\cite{Saito-etal:77}, while the profile of \cite{Leblanc-etal:98} gives
densities too low by a factor $\approx 3$.

  The agreement with  the temperature profile given by the ratio \ion{Mg}{x}
  62.5~nm/ \ion{Mg}{ix} 36.8~nm, while still good, is somewhat less stringent,
  with a coronal temperature of about $1.5 - 1.6\; 10^6$~K, slightly higher than
  the peak value of $1.45\times 10^6$~K given by both \cite{Gibson-etal:99} and
  \cite{Vasquez-etal:03}.  

\subsubsection{Polar Regions (Coronal Holes)}

In polar regions, the inferred densities match much better the estimates by
\cite{Kohl-etal:98} than those by the other sources:
\cite{Guhathakurta-etal:99,Munro-Jackson:77,Sittler-Guhathakurta:99,Sittler-Guhathakurta:02},
and \cite{Saito-etal:77}.  This
last article gives a density $N_\mathrm{e} \approx 2\times 10^6$~cm$^{-3}$ at
$r/R_\odot=1.2$, below the lower limit of Figure \ref{fig:cds_ratios_ch}.

  The agreement of the temperature from the \ion{Mg}{x} 62.5~nm/\ion{Mg}{ix}
  36.8~nm ratio with the profile given by \cite{Vasquez-etal:03} is in this
  case excellent, and much better than with the profile of
  \cite{Guhathakurta-etal:99}.

\section{Calculated Radiances}\label{sec:calcs}

The calculated radiances can be divided into a collisional component and a
component radiatively excited by the line radiation from the solar disk (for
the \ion{He}{ii} line or other transition-region lines) or from the inner
corona (for the \ion{Si}{xi} line and other coronal lines).  As an additional
consistency check in the calculations, we applied the procedure described
below to two other lines observed by CDS: 
\ion{Si}{xii} 52.1~nm and \ion{Mg}{x} 62.5~nm.

  The collisional component of a line emissivity for the strong lines we are
  considering can be computed using the standard approximation of 
  collisional excitation from the ground level followed by radiative decay.

  The radiative component of the emissivity due to resonant scattering, \textit{i.e.}
  the line emission due to the absorption and re-emission of photons from the
  lower solar atmosphere (chromosphere or near-limb corona) by ions/atoms in
  the corona (also known as ``Doppler dimming''), has been calculated adopting
  the formalism of \cite{Withbroe-etal:82} and \cite{Noci-etal:87}.
  The relevant atomic data adopted were taken from the CHIANTI database
  version 6 \citep{Dere-etal:97,Dere-etal:09}.

Finally, the integration of line emissivities
is performed along the line of sight (\,LOS\,) over a coronal
sector $\approx 120^{\circ}$ wide, where a spherical symmetry is assumed. At low
heights the most significant contribution to the synthesized line radiance is
given by the emission near the plane of the sky (\,POS\,), because of the rapid
decrease of the electron density with increasing heliocentric distance. For
larger heliocentric heights, the integration path along the LOS must be
larger since the electron density decreases more slowly; therefore, even the
emission far from the POS contributes significantly to the line intensity. For
all the heliocentric heights where the coronal-line radiances are assessed, a
further increase of the integration interval involves a radiance variation
$<5\%$.

\subsection{Coronal Physical Quantities}\label{sec:calcs:corona}

  The other ingredients needed to compute line emissivities are as follows: electron
  density and temperature (summarized in
  Figure \ref{fig:electron_temperature_density}), ion abundance, outflow speed
  of the expanding solar corona, three-dimensional (\,3D\,) velocity distribution of the
  coronal absorbing ions as a function of heliocentric distance and latitude, 
  total radiance and profile of the exciting lines emitted by the
  disk or the lower corona.  These parameters will be discussed in more
  detail in the following sections.

All of the physical and dynamic parameters describing the line-profile formation
along the LOS are chosen by employing coronal models consistent with the
observations near solar minimum, and already briefly discussed in the
  previous section in the context of empirical diagnostics. In some cases, these models are extrapolated down to $1.2\,R_{\odot}$
to reach the SOHO/CDS domain of validity and at high heliocentric distances in
the coronal sector $\approx 120^{\circ}$ wide where the integration along the LOS
is performed. 
\subsubsection{Electron Temperature and Density: Polar Regions}\label{sec:calcs:corona:TP:ch}
Electron densities can be determined from measurements of polarized brightness
(\,pB\,) by using the inversion technique developed by \cite{vandeHulst:50}. 
The pB is directly related to the coronal electron density,
since it depends upon the Thomson scattering of the photospheric white-light
radiation by coronal electrons. The technique relies on the inversion of the
equation describing the relation between the observed Thomson-scattered
pB and the electron density.

Among the first attempts to infer the electron-density distribution within a
polar coronal hole from pB observations with space coronagraphs,
\cite{Munro-Jackson:77} and \cite{Saito-etal:77} derived electron-density
radial profiles, valid from 2 to $5\,R_{\odot}$ and from 2.5 to
$5.5\,R_{\odot}$, respectively, by using data provided by the High Altitude
Observatory (\,HAO\,) white-light
coronagraph on \textit{Skylab}.

More recently, \cite{Kohl-etal:98} modeled the electron density in a polar
coronal hole, from $\approx 2$ to $4\,R_{\odot}$, by measuring the linear
polarization due to Thomson-scattered photospheric light with the 
SOHO/\textit{Ultraviolet Coronagraph Spectrometer} \citep[UVCS:][]{Kohl-etal:1995}
white-light channel, while \cite{Sittler-Guhathakurta:99} developed an
empirical electron-density profile at the poles from \textit{Skylab} coronagraph
white-light data \citep{Guhathakurta-etal:96}. The electron-density model,
valid only outside $1.16\,R_{\odot}$, was extended into interplanetary space
by using electron densities derived from the \textit{Ulysses} plasma data
\citep{Phillips-etal:95}.

Quantitative information on the electron-density distribution of a coronal hole
was also estimated by \cite{Guhathakurta-etal:99} by combining white-light,
pB observations from the SOHO/LASCO-C2 and -C3 and HAO/Mauna
Loa Mark\,\textsc{III} coronagraphs with density-sensitive EUV line ratios of
\ion{Si}{ix} 35.0/ 34.2 nm observed by SOHO/CDS, to obtain a density
profile from 1 to $8\,R_{\odot}$ for the polar coronal hole. With the
assumptions that the coronal gas is locally isothermal and in radial
hydrostatic equilibrium along the LOS, \cite{Guhathakurta-etal:99} determined
an effective (electron) temperature in polar coronal holes.

Another method for inferring the coronal electron temperature is based on \textit{in-situ} 
measurements of the solar-wind charge state, which is determined in large
part by the electron temperature in the inner corona, where the ionization and
recombination times are still short compared to the solar-wind expansion
time. The electron temperature between 1 and
$8\,R_{\odot}$ derived  by
\cite{Ko-etal:97} and \cite{Cranmer-etal:99} in a polar coronal hole, from
observations with \textit{Ulysses}/SWICS, were fitted with a combination of two power
laws from \cite{Vasquez-etal:03} to give an electron-temperature model at the
poles.

The above models, which differ significantly for $r<2\,R_{\odot}$ as shown in
the left panels of Figure \ref{fig:electron_temperature_density}, are employed
in the synthesis of the radiance of the \ion{He}{ii} 30.4~nm,
\ion{Si}{xi} 30.3~nm, \ion{Si}{xii} 52.1~nm, and \ion{Mg}{x}
62.5~nm lines, in order to point out the sensitivity of the coronal line
intensity to the density and especially to the temperature of the electrons.

\subsubsection{Electron Temperature and Density: Equatorial Regions}\label{sec:calcs:corona:TP:eq}
Concerning the electron-density models employed in the assessment of
the intensity of the \ion{He}{ii}, \ion{Si}{xi}, \ion{Si}{xii}, and
\ion{Mg}{x} lines at low heliographic latitudes, \cite{Saito-etal:77} and
\cite{Sittler-Guhathakurta:99} (see
  Section \ref{sec:usun:diagnostics})
also  investigated the equatorial electron-density distribution, providing
models valid from 2.5 to $5.5\,R_{\odot}$ and outside $1.16\,R_{\odot}$,
respectively. The electron-density models developed by
\cite{Leblanc-etal:98}, \cite{Gibson-etal:99}, and \cite{Thernisien-Howard:06}
within a streamer have been considered in the analysis as well.

\cite{Leblanc-etal:98} derived the electron-density distribution in the
ecliptic plane, from the corona to $1$~AU, using observations from
$13.8$~MHz to a few kilohertz by the radio experiment WAVES onboard the \textit{Wind}
spacecraft. The radio technique is based on the measurement of the drift rates of
type\,\textsc{III} bursts at each frequency, which allowed the authors to
estimate the speed of electron streams creating the bursts, and, indirectly,
an electron-density model all along the trajectory of the bursts for
$r>1.2\,R_{\odot}$.

\cite{Gibson-etal:99} modeled electron densities within a streamer structure,
from 1 to $8\,R_{\odot}$, using coronal observations of both visible white
light (SOHO/LASCO-C2 and HAO/Mauna Loa Mark\,\textsc{III} coronagraphs) and
extreme ultraviolet (SOHO/CDS spectrometer) emission, by way of the van de
Hulst inversion.

Finally, \cite{Thernisien-Howard:06} presented a 3D reconstruction of the
electron density of a streamer, using total-brightness observations performed
by SOHO/LASCO-C2 and -C3 from 2.5 to $30\,R_{\odot}$. The radial profile of the
electron density was determined via inversion techniques based on the method
implemented by \cite{Hayes-etal:01}.

The above electron-density models, which agree quite well for
$r<1.5\,R_{\odot}$ while differing significantly for larger heliocentric
distances, are shown in the bottom-right panel of
Figure \ref{fig:electron_temperature_density}. In the top-right panel of the
same figure, the electron-temperature models within an equatorial streamer by
\cite{Gibson-etal:99} and \cite{Vasquez-etal:03} are shown.

From the inferred density profile, \cite{Gibson-etal:99} deduced a
hydrostatic temperature profile inside the streamer from the ideal gas law,
$T=(1+2\alpha)/(2+3\alpha)\times P/(n_\mathrm{e}k_\mathrm{B})$. Here
$\alpha=0.1$ is the helium number density relative to hydrogen and
$P$ is the total thermal pressure, which, in the assumption that the plasma in
the streamer is in hydrostatic equilibrium, is exactly balanced by gravity.

Because of the high density and low outflow velocity in streamers, 
protons and electrons are expected to be in thermal equilibrium with each
other, so the electron temperature at the Equator can be assumed to be equal
to the proton temperature. Furthermore, because of the rapid charge exchange
between protons and neutral-hydrogen atoms, the hydrogen temperature, which
can be estimated by measuring the width of the Ly\,$\alpha$ coronal
emission line profile, reflects the proton temperature at radii up to
$3\,R_{\odot}$ in the polar regions and even higher in the equatorial regions
\citep{Withbroe-etal:82}. Hence, \cite{Vasquez-etal:03} derived an
electron-temperature 
model within a streamer, by fitting (with a combination of two
power laws) SOHO/UVCS low-latitude observations of the Ly\,$\alpha$
line width \citep{Raymond-etal:97}.
\subsubsection{Elemental Abundances}\label{sec:calcs:corona:abundances}
Because of the numerous of results on elemental abundances derived
from various types of data provided by different instruments and under
different physical conditions (solar energetic particle (\,SEP\,) observations, flare observations,
spectroscopy of closed coronal loops, corotating interaction region analysis),
it is often quite difficult to know which data set
should be used to ensure consistency with the observations.

In this work, we adopted the set of photospheric abundances listed in
  \cite{Asplund-etal:09}; more specifically,
$A_\mathrm{He}=0.085$, $A_\mathrm{Si}=3.24\times10^{-5}$, and
$A_\mathrm{Mg}=3.98\times10^{-5}$. 
To a first approximation, no variation of the elemental abundances
with heliographic latitude and heliocentric distance is considered here; \textit{i.e.}
the elemental abundances are assumed to be equal in the
polar and equatorial regions and to be constant throughout the analyzed
height range.
  However, some observations indicate that equatorial regions at the solar
  minimum include a ``coronal'' mix of elements, \textit{i.e.} elements of low FIP
  such as Si and Mg should be more abundant by a
  factor of several.  

  This dichotomy between abundances in the equatorial and polar regions is
  well established from \textit{in-situ} measurements (\textit{e.g.},
  \citealt{vonSteiger-etal:00}, who give a FIP bias of two\,--\,three), but is probably a
  rather coarse schematization for the corona closer to the Sun
  \citep[\textit{e.g.},][]{Raymond-etal:97}, a region for which there are
  various contradictory measurements.
  We mention, for instance, \cite{Feldman-etal:98} who considered
  abundances in the low corona above polar coronal holes and the quiet Sun
  during minimum.  They measured a FIP enhancement above the quiet Sun of
  about four between, \textit{e.g.}, Mg and Ne.
  On the other hand, \cite{Young:05} instead found the Mg/Ne ratio 
  to be nearly photospheric.
  Moreover, the He abundance is
  rather variable even in the solar wind; for instance, there are 
  measurements indicating $A_\mathrm{He}$ of the order of $0.05$ in the fast
  solar wind \citep[\textit{e.g.},][]{Bochsler:98}.  

  In any case, the line radiances
  scale linearly with the elemental abundance, so that it is easy to evaluate the
  effect of a given chemical mixture on the calculations discussed in this article.
\subsubsection{Outflow Speed of the Expanding Coronal Plasma}\label{sec:calcs:corona:flow}
  In order to quantify the Doppler dimming of the resonant scattering process
  due to the expansion of the solar corona, 
  a
  semiempirical model for the outflow speed of the fast solar wind has been
  employed in the analysis. 

  The model
has been derived so as to be compatible
with the previous results by \cite{Antonucci-etal:04} at low heliocentric
distances and with the asymptotic outflow velocity
$v_{\infty}\approx 750-800$ \kms, which is definitely achieved at
$\approx 20\,R_{\odot}$ \citep{Breen-etal:96}, but probably even at
$\approx 8\,R_{\odot}$, according to interplanetary-scintillation observations
\citep{Grall-etal:95}. In particular, the model is consistent with the outflow
velocity of the oxygen component of the fast solar wind derived by
\cite{Telloni-etal:07a,Telloni-etal:07b}, from the intensity ratio of the
Doppler dimmed \ion{O}{vi} 103.19, 103.76~nm doublet lines observed
with SOHO/UVCS in polar coronal holes, from 1.5 out to $5\,R_{\odot}$, during
the 1996\,--\,1997 solar minimum.

Concerning the flow speed in the streamer regions,
\cite{Sheeley-etal:97} developed a model for the slow solar wind from 2 to
$30\,R_{\odot}$, by tracking the birth and outflow of outward moving density
inhomogeneities observed with the SOHO/LASCO-C2 and -C3 coronagraphs during
sunspot-minimum conditions in 1996. The authors concluded that those
coronal moving features were passively tracing the slow wind, which originates
above the cusps of helmet streamers at about $3-4\,R_{\odot}$ and radially
outflows with a nearly constant acceleration of about $4$~m s$^{-2}$,
according to a parabolic speed profile. This profile is consistent with an
isothermal solar wind expansion at a temperature of about $1.1$~MK and a
sonic point near $5\,R_{\odot}$. This model is also
consistent with the results found by \cite{Antonucci-etal:05} from SOHO/UVCS
data, in the analysis of the slow wind and magnetic topology in the solar
minimum corona in 1996\,--\,1997.
\subsubsection{Intensity Profiles from the Chromosphere and Lower Corona}\label{sec:calcs:corona:pumping}

    In order to estimate the optical pumping by disk radiation of the lines 
    considered in this article, we use the average USUN CDS line profiles; for
    the \ion{He}{ii} 30.4 and \ion{Si}{xi} 30.3 nm lines the average disk line
    radiances are $5.39\times 10^3$ and $1.04\times 10^2$ erg cm$^{-2}$
    s$^{-1}$ sr$^{-1}$, respectively, while for the \ion{Mg}{x} 62.5 and
    \ion{Si}{xii} 52.1 nm lines the measured average radiances are $50.4$ and $3.81$
    erg cm$^{-2}$ s$^{-1}$ sr$^{-1}$, respectively.  We remark again that
    the \ion{Si}{xi} 30.3 nm disk radiance has been taken into account for the
    calculation of the pumping of both the \ion{Si}{xi} and the
    \ion{He}{ii} lines in the corona.

In all cases, a constant intensity and shape of the
  exciting line profiles across the solar disk is assumed, \textit{i.e.} no limb
brightening or darkening has been considered.

\subsubsection{3D Velocity Distribution of the Coronal Helium Ions}\label{sec:calcs:corona:anisotropy}
In order to quantify the number of chromospheric photons scattered by the
He$^{+}$ ions in the extended corona, we account for the width of the coronal
absorption profile of the helium ions along the direction of the incident
radiation $[\textbf{n}]$, which is related to the helium kinetic temperature
$[T_{\mathrm{He},n}]$. The velocity distribution of the absorbing helium ions
sets the assumed degree of anisotropy. In particular, if
$T_{\mathrm{He},n}=T_\mathrm{e}$ the degree of anisotropy is maximum, while if
$T_{\mathrm{He},n}=T_\mathrm{He,LOS}$, where $T_\mathrm{He,LOS}$ is the helium
temperature along the LOS, the ion velocity distribution is isotropic. 
There is some uncertainty related to the temperature anisotropy
assumed in the analysis. Moreover, further uncertainty is introduced since the
helium temperature along the LOS has not yet been directly measured. In the
synthesis of the radiance of the \ion{He}{ii} 30.4~nm line, it is assumed
that the helium temperature along the LOS reflects that of the hydrogen atoms,
both in polar and equatorial regions. In particular, a semiempirical model of
the helium temperature has been derived from the hydrogen temperature inferred
by measuring the width of the Ly\,$\alpha$ coronal emission line profile
observed with the SOHO/UVCS spectrometer during the 1996\,--\,1997 minimum
\citep{Antonucci-etal:05}. Since there is no evidence for anisotropic
velocities of helium ions, we assume here that they have an isotropic
velocity distribution both at high and low heliographic latitudes. 
Note that the above assumptions do not affect significantly the \ion{He}{ii}
intensity results for heliocentric distances smaller than $2\,R_{\odot}$,
where the collision excitation dominates, whereas at larger heliocentric
distances they lead to uncertainties of less than 30\%.

\section{Discussion}\label{sec:comparison}

\begin{figure}[b!]
  \centering
  \includegraphics[trim=40 30 40 20,clip=false,width=0.95\linewidth]%
  {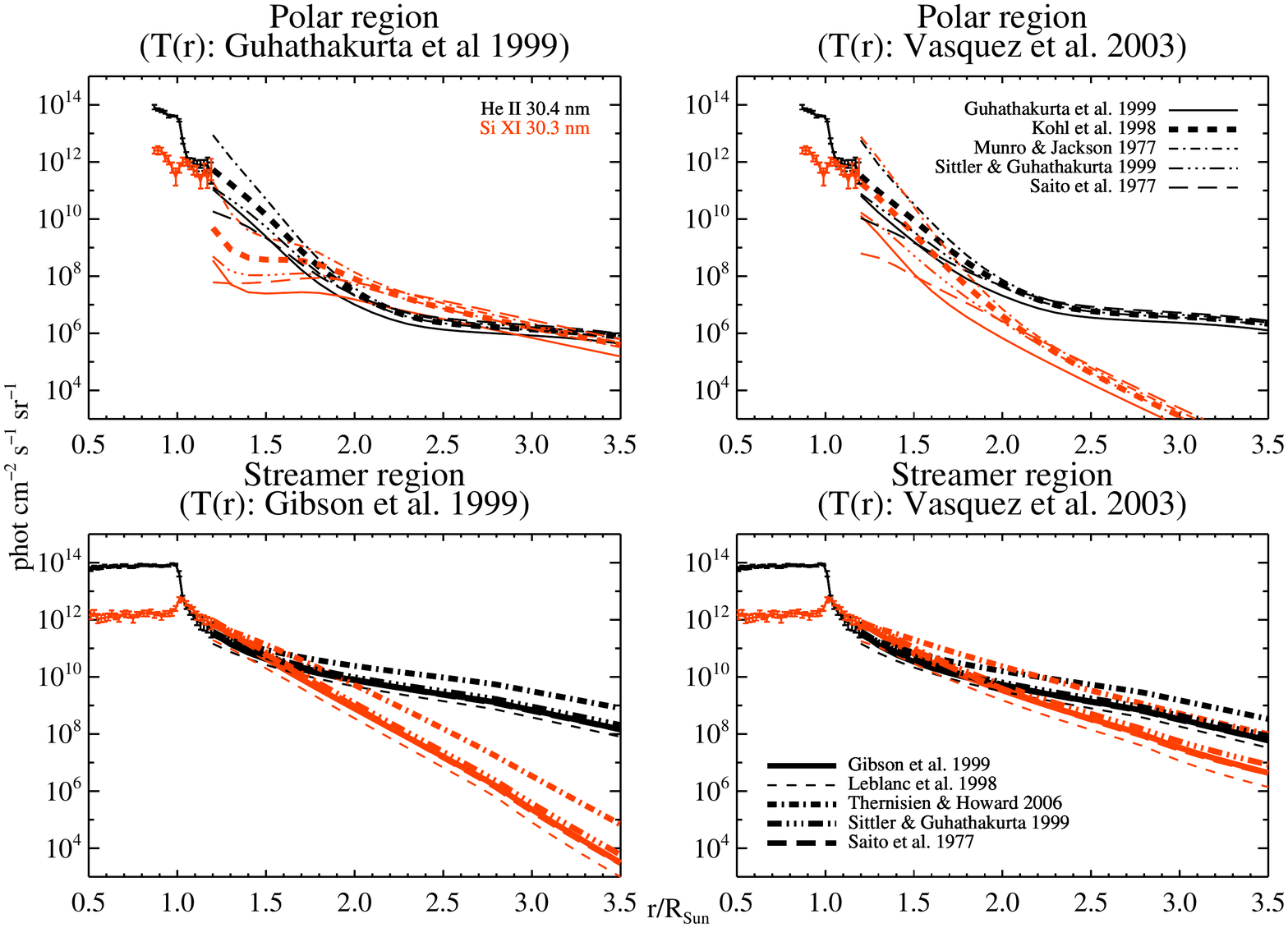}%
  \caption{%
    Comparison of radiances measured from SOHO/CDS and computed adopting
    various model profiles of temperature and density in the corona.  The
    lines shown here are: \ion{He}{ii} 30.4~nm (black) and 
    \protect\ion{Si}{xi} 30.3~nm (red).
  }
  \label{fig:compare_radiances_1}
\end{figure}
\begin{figure}[b!]
  \centering
  \includegraphics[trim=40 30 40 20,clip=false,width=0.95\linewidth]%
  {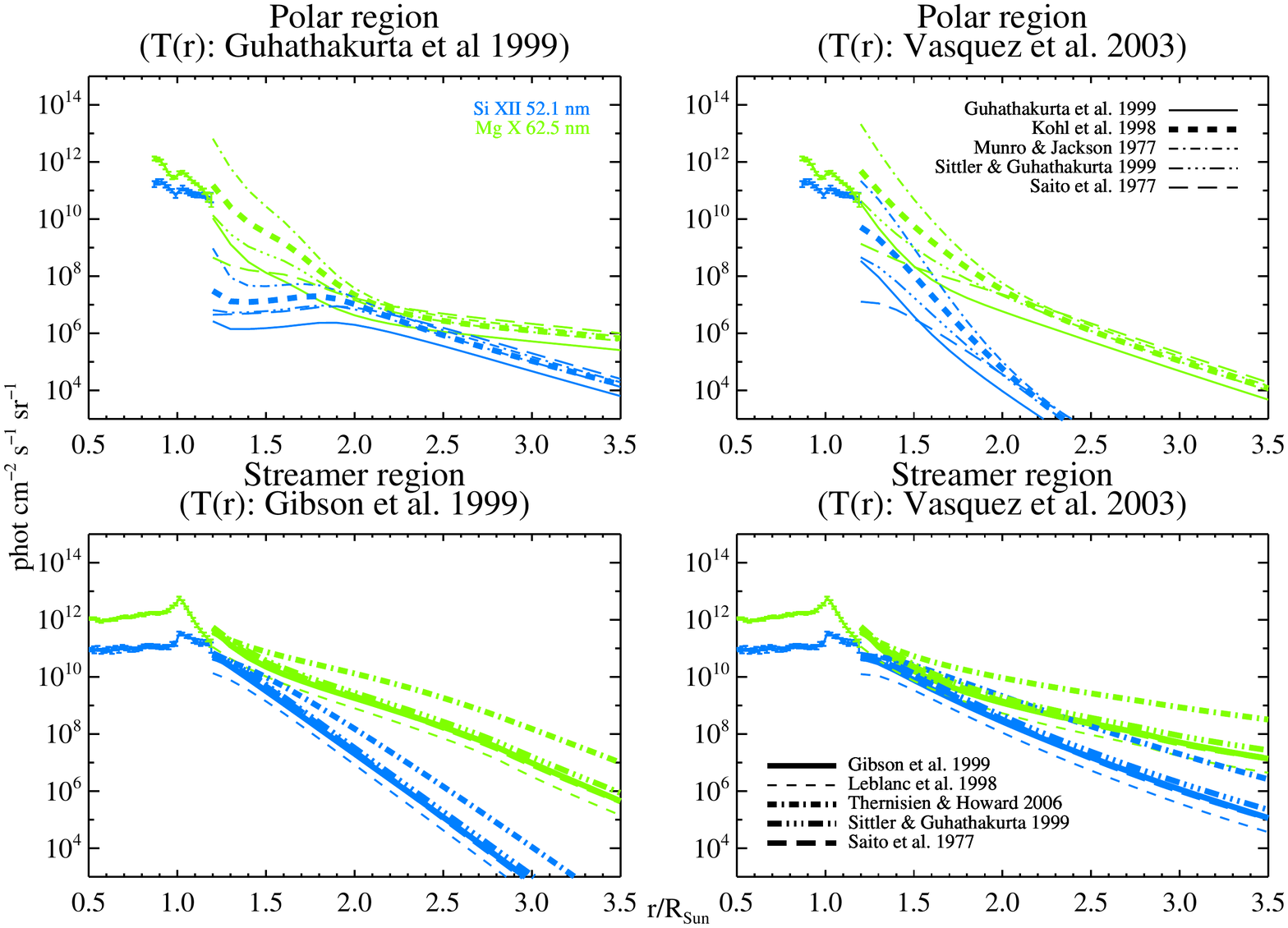}
  \caption{%
    Comparison of radiances measured from SOHO/CDS and computed adopting
    various model profiles of temperature and density in the corona.  The
    lines shown here are: \ion{Si}{xii} 52.1~nm (blue) and 
    \protect\ion{Mg}{x} 62.5~nm (green).
  }
  \label{fig:compare_radiances_2}
\end{figure}

The radiances computed as described in Section \ref{sec:calcs} can be
compared with the radiances measured as described in Section \ref{sec:usun}, in
the region $r/R_\odot\approx 1.2$, delimiting the respective domains of validity.
We again remark that while some coronal temperature and density
  models are already given down to that value of heliocentric distance
  \citep[as in, \textit{e.g.},][]{Gibson-etal:99}, in other cases the values shown here
  are extrapolated below their domain of validity \cite[as it is the case of
  data from, \textit{e.g.},][]{Thernisien-Howard:06}, and therefore should be treated
  with caution (see also Figure \ref{fig:electron_temperature_density}).

The comparison for the lines \ion{He}{ii} 30.4~nm and \ion{Si}{xi} 30.3~nm is
shown in Figure \ref{fig:compare_radiances_1}; for that comparison, 
note that measured \ion{He}{ii} 30.4~nm radiances are dominated
by stray light almost everywhere above the limb, and thus constitute
only an upper limit for the theoretical calculations.
Figure \ref{fig:compare_radiances_2} shows instead the measured and computed
radiances of the \ion{Si}{xii} 52.1~nm and \ion{Mg}{x} 62.5~nm lines.  As
noted in Section \ref{sec:usun:diagnostics}, the former line is also
likely to be affected to some extent by instrumental stray light, especially
in the corona above polar holes; in those cases, its measured radiances can be
regarded as upper limits.

  Note that we have also examined the relative
  weight of the collisional and radiative components in the various computed
  lines.  Normally, it is assumed that collisional excitation dominates in the
  low corona for most ions/atoms, while further away from the Sun both
  radiative and collisional excitation may be important. In practice, we found
  that the \ion{Si}{xii} 52.1~nm line is practically purely
  collisional, while for the \ion{Si}{xi}~30.3 nm and \ion{Mg}{x} 62.5~nm
  lines the radiative contribution is not always completely negligible, but is
  at most of the order of 30\,--\,40\% (at $\approx 3\, r/R_\odot$), depending on the
  specific set of approximations used \citep[see also][]{Withbroe-etal:82}.
  However, the \ion{He}{ii} 30.4~nm line has a negligible
  collisional component and can thus be considered purely radiative, and
  therefore dependent on the details of the flow velocity in the corona, among
  other things (Doppler dimming).

\subsection{Streamer}\label{sec:comparison:eq}

The comparison between computed and measured radiances in the streamer region
of the data set considered here is shown in the lower panels of
Figures \ref{fig:compare_radiances_1} and \ref{fig:compare_radiances_2}.  Taking
into account that the radiances of the \ion{He}{ii} 30.4~nm line above the
limb are strongly affected by stray light, all calculations appear to
reasonably match the observations.  In particular, both the temperature
profiles of \cite{Gibson-etal:99} and \cite{Vasquez-etal:03} reproduce the
radiances in the corona observed by CDS.  This is not entirely surprising,
since both temperature profiles set their peak coronal temperature at $\approx
1.5\times 10^6$~K, a value close to the one inferred from CDS line ratios (see
Figure \ref{fig:cds_ratios_eq}).  
Moreover, almost all of the density profiles
considered here produce radiances that can be considered in accordance with
CDS observations.  Nevertheless, in the following, on the basis of the
analysis of Section \ref{sec:usun:diagnostics}, we will exclude the calculations
obtained with the density profile of \cite{Leblanc-etal:98} (short-dashed
lines in the figures).

  Overall, the agreement between computed and measured radiances is
  considered reasonable, but the computed \ion{Mg}{x} 62.5~nm seems to be a
  factor $\approx$~two higher than the measurements.  A better agreement would
  be found if the density profile of \cite{Leblanc-etal:98} were adopted, but that
  density profile is incompatible with the densities inferred from
  the \ion{Si}{ix} line ratio, as mentioned above.

  The problem with the \ion{Mg}{x} 62.5~nm line is not completely unexpected:
  discrepancies between observed and predicted radiances/irradiances (assuming ionization
  equilibrium and some form of emission-measure modeling)
  in lines from the Na- and Li-like ions have 
  appeared since the beginning of space observations.  This has been noted
  occasionally in the literature.  The problem is actually more complex and
  pervasive than originally understood, as explained in \cite{DelZanna:99}.

  \cite{DelZanna:99} found with differential emission measure (\,DEM\,) analysis that \ion{Mg}{x} lines in the
  quiet Sun were overestimated by a factor of 1.6.  \cite{DelZanna:99}
  reconsidered various historical records \citep[\textit{e.g.},][]{Judge-etal:95} and
  found that the problem is present in all data sets.  The lines from Li-like
  ions formed in the transition region tend to be underestimated by a large
  factor, while those formed at coronal temperatures are overestimated.
  Finally, \cite{DelZanna-etal:01b} reanalyzed \textit{Skylab} data and found that
  \ion{Mg}{x} lines were overestimated by a factor of ten.

  For a short review on the subject, see \cite{delzanna_etal:02_aumic}.
    We only mention here that the problem can perhaps be ascribed to
    the ion-balance calculations, although some of the explanations proposed
    (\textit{e.g.} the density dependence of dielectronic recombination rates)
    probably do not apply to these coronal observations.

\subsection{Polar Regions (Coronal Holes)}\label{sec:comparison:ch}

From the upper panels of Figures \ref{fig:compare_radiances_1} and
\ref{fig:compare_radiances_2}, it is clear that \ion{Si}{xi} 30.3~nm and
\ion{Si}{xii} 52.1~nm radiances computed with the temperature profile of
\cite{Guhathakurta-etal:99} strongly underestimate
(even by two or more orders of
magnitude) the observed values in the lower corona, regardless of the density
profile adopted.  This is an independent confirmation of the same finding
discussed in Section \ref{sec:usun:diagnostics} (see also
Figure \ref{fig:cds_ratios_ch}).  
Also considering the upper limits
given by the \ion{He}{ii} 30.4~nm line, and the radiances of the \ion{Mg}{x}
62.5~nm line, the radiance profiles that best match the observations are
instead those obtained with the \cite{Vasquez-etal:03} temperatures and the
\cite{Kohl-etal:98} densities. As discussed in
Section \ref{sec:usun:diagnostics}, these are also the $T_\mathrm{e}(r)$ and
$N_\mathrm{e}(r)$ profiles, respectively, that best match the line ratios in
Figures \ref{fig:cds_ratios_eq} and \ref{fig:cds_ratios_ch}.  

\subsection{The Ratio \ion{Si}{xi} 30.3~nm/\ion{He}{ii} 30.4~nm}\label{sec:comparison:ratio}

\begin{figure}[htb]
  \centering
  \includegraphics[trim=40 30 40 20,clip=false,width=0.95\linewidth]%
  {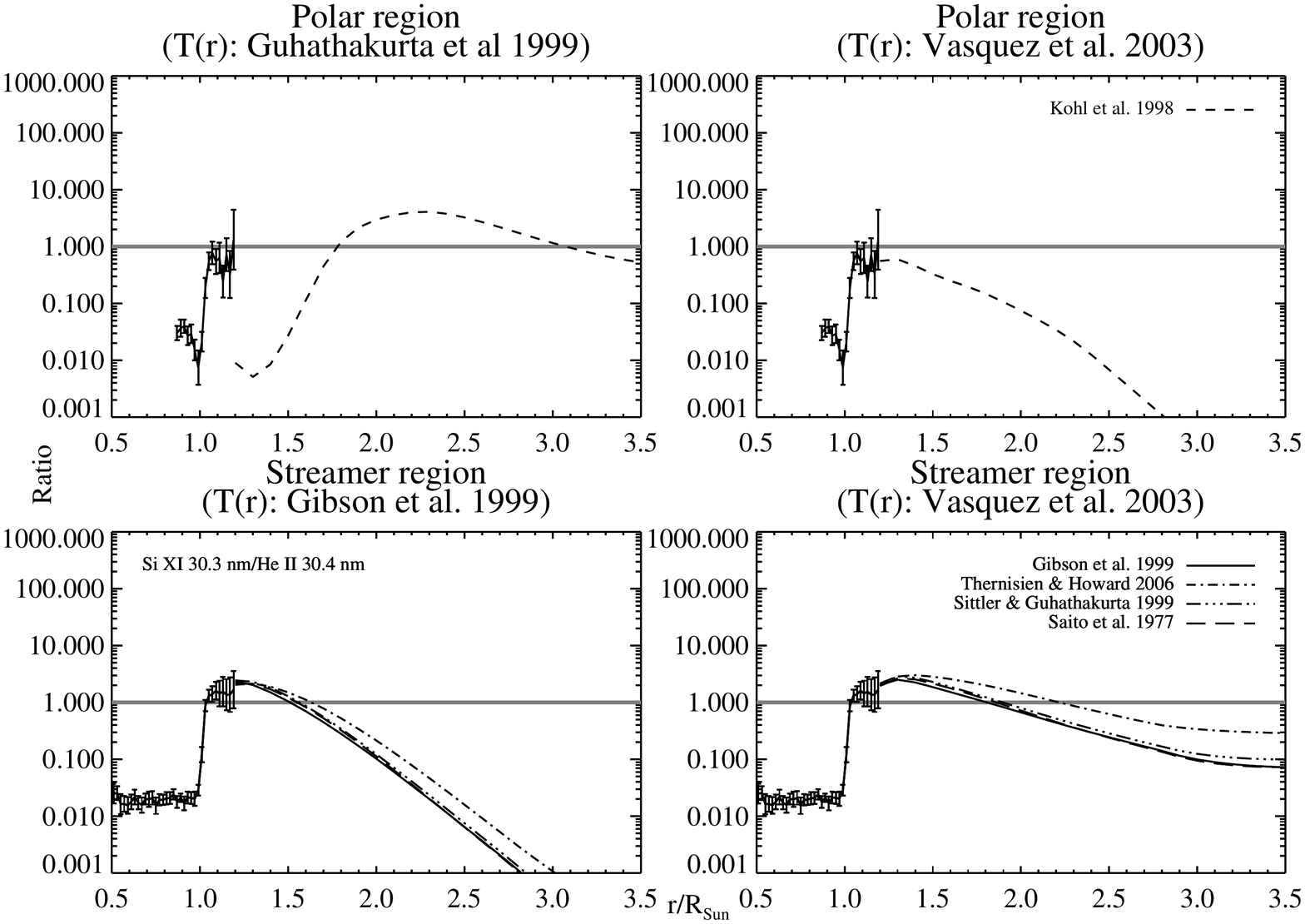}
  \caption{%
    Ratio \protect\ion{Si}{xi} 30.3~nm/\protect\ion{He}{ii} 30.4~nm, both
    observed by CDS ($r/R_\odot<1.2$) and computed ($r/R_\odot>1.2$). The
    horizontal grey line indicates, for clarity, the level at which the two
    lines have the same radiance.
  }
  \label{fig:compare_ratio}
\end{figure}

From the measured and computed values of the radiances of the \ion{Si}{xi}
30.3~nm and \ion{He}{ii} 30.4~nm lines, we can then proceed to determine
the range of values of their ratios in the quiescent solar corona.  In
Figure \ref{fig:compare_ratio} we show the observed ratio from the CDS
observations, up to $1.2\; R_\odot$.  Recalling once again that the \ion{He}{ii} 30.4~nm
line is affected by stray light above the limb, the
corresponding values of the ratio shown in that figure are actually only upper
limits.

For the polar regions, we show the ratio computed with the temperatures
of \cite{Vasquez-etal:03} and the densities of \cite{Kohl-etal:98} (upper
right panel of Figure \ref{fig:compare_ratio}), because those profiles produce a
better match with observations, as discussed in the previous section.  The
result is that the helium Ly\,$\alpha$ dominates the
nearby \ion{Si}{xi} resonance line even in the lower corona, down to
  $\approx 1.2\; R_\odot$.  However, we also show the results
obtained with the temperature profile of \cite{Guhathakurta-etal:99} (upper
left panel of Figure \ref{fig:compare_ratio}), even though it clearly
underestimates the radiances of several coronal lines in the lower corona, to
show the effect of the temperature profile on these results.

In the case of a streamer above $\approx 1.5\; R_\odot$, the temperature profile
also has a strong effect, especially on the \ion{Si}{xi} 30.3~nm line (see
Figure \ref{fig:compare_radiances_1}).  On the other hand, the details of the
density profile are much less important in determining the value of the ratio
\ion{Si}{xi} 30.3~nm/\ion{He}{ii} 30.4~nm.  However, in all cases the
\ion{Si}{xi} 30.3~nm line dominates over the \ion{He}{ii} 30.4~nm line in the
corona below $\approx 1.5\; R_\odot$, and even above (at least up to $\approx 2.3\;
R_\odot$) if we adopt the temperature profile of \cite{Vasquez-etal:03}.

  As mentioned in Section \ref{sec:calcs:corona}, the above results are based on
  the assumption of a photospheric mixture of elements.  

  In polar holes, the Si abundance should not be different, but He might be
  depleted by a factor $\approx$~two with respect to the photosphere.  If so, the
  ratio \ion{Si}{xi} 30.3~nm/\ion{He}{ii} 30.4~nm could increase above unity
  for heliocentric distances below $\approx 1.5\; R_\odot$ (considering only the
  temperature profile of \citealt{Vasquez-etal:03}).

  However, in quiescent equatorial streamers it is possible that
  the abundance of a low-FIP element like Si is enhanced with respect to a
  high-FIP element like He, by a factor of the order of three\,--\,four.  Such an
  enhancement would extend considerably the region where the \ion{Si}{xi}
  30.3~nm dominates the \ion{He}{ii} 30.4~nm line. But
  the agreement with the radiances shown in Figure \ref{fig:compare_radiances_1}
  would also be less satisfactory: increasing the abundance in streamers of
  the low-FIP elements Si and Mg would perhaps improve the agreement for the
  \ion{Si}{xii} 52.1 line, at the expense of a worse match with observations
  for the \ion{Si}{xi} 30.3 and \ion{Mg}{x} 62.5 lines.  Clearly, in this case
  invoking for example a ``filling factor'' would not mitigate the
  problem.

\section{Conclusions}\label{sec:conclusions}

Our results indicate that the \ion{Si}{xi}
30.3~nm line is important compared to the \ion{He}{ii} 30.4~nm in
the corona below $\approx 2.0\; R_\odot$, to the point of being the dominant
source of emission in the 30.4~nm band in streamers, less likely so
  in polar regions (coronal holes).  Beyond that distance, different
temperature profiles predict a generally rapid decline of that
contribution, mainly because of the strong dependence of the \ion{Si}{xi}
resonance line on temperature.  For instance, the model by \cite{Vasquez-etal:03}
predicts a non-negligible contribution of the \ion{Si}{xi} 30.3~nm
up to $\approx 3.0,\, 3.5\; R_\odot$, whereas the temperature profile from
\cite{Gibson-etal:99} implies a much more rapid drop of that line with
heliocentric distance.

However, the density profile has a relatively less
important role in determining the emission ratio \ion{Si}{xi}/\ion{He}{ii} in
the extended corona.

There is one important caveat about the \ion{Si}{xi} 30.3~nm 
calculations. The CHIANTI  atomic data include 
rates for collisional excitation by electrons that were not
calculated, but 
interpolated along the Be-like sequence, 
and the same occurred for \ion{Mg}{ix}. 
\cite{delzanna_etal:08_mg_9} performed the first 
scattering calculation for \ion{Mg}{ix}, and found significant 
differences in the emissivities of key lines
 when compared to the results from the interpolated data.
The resonance line was only affected by about 10\%, 
so large corrections to the \ion{Si}{xi} 30.3~nm are not 
expected. However, we will know this for sure only when 
new calculations (in progress by GDZ) become available.

  As a consistency check for our calculations, we also computed the radiances
  for a few representative lines observed by CDS, taking into account both the
  collisional excitation and the radiative pumping by the disk radiation,
  using average radiances from the same CDS data.  Here we only
  show the results for the \ion{Mg}{x} 62.5~nm and \ion{Si}{xii} 52.1~nm
  lines.  The overall agreement between calculations and observations is
  reasonable; there are some discrepancies in the case of the \ion{Mg}{x}
  line, but they can be attributed to a known problem (but of
  uncertain origin) in the emissivity calculations
  for that line.

We recall that the observed radiances shown here were derived adopting the
latest radiometric calibration for the CDS/NIS, including the correction for
the long-term decay in sensitivity of the instrument \citep[and references
therein]{DelZanna-etal:10}.  Since most, if not all, previous analyses done
merging near-limb EUV data and extended corona measurements rely on line
ratios only \citep[\textit{e.g.},][]{Gibson-etal:99}, we note the
generally good to excellent match between observed and computed
\emph{absolute} radiances in the strongest coronal lines.

  Finally, our results were obtained adopting photospheric abundances from
  \cite{Asplund-etal:09}.  Adopting coronal abundances
  would alter somewhat the results, especially for $r
  < 2.0\; R_\odot$.  However, in that region abundance measurements are less
  well established than, for instance, \textit{in-situ} wind measurements: the
  effect of abundance variations should thus be considered as an
  additional source of uncertainty in the ratio \ion{Si}{xi}
  30.3~nm/\ion{He}{ii} 30.4~nm.

In summary, it is true that the diagnostic potential of narrowband imaging
around 30.4~nm in the extended corona is high.  In particular, radiative
excitation of the main line in the band, \ion{He}{ii} 30.4~nm, from disk
emission of either the same line or of the nearby \ion{Si}{xi} 30.3~nm line,
can provide a useful tool for diagnosing radial velocities.  However, the full
diagnostic potential of this band can only be fulfilled if it is possible to
place constraints on the possible ``contamination'' from coronal \ion{Si}{xi}
30.3~nm emission.  In particular, we have shown that the ratio \ion{Si}{xi}
30.3~nm/\ion{He}{ii} 30.4~nm is very sensitive to the specific temperature
profile adopted for the corona, which is normally rather uncertain.

A corollary of these calculations is that the diagnostic potential of
narrowband observations in the 30.4~nm band in critical regions of the corona
($<3\, R_\odot$) can only be fully exploited if aided by spectroscopic
observations capable of disentangling the \ion{Si}{xi} contribution from the
\ion{He}{ii} emission.  Alternatively, if 
one could find a way to independently determine
the coronal temperature profile (the density profile is much less critical in
this respect), perhaps it could be aided through simultaneous visible light imaging.
Both approaches will be feasible with METIS 
onboard \textit{Solar Orbiter}.

\begin{acks}
  This work is part of the scientific activities supporting the
    development of METIS, and as such has been supported in part (Italy) by
    ASI-INAF contract I/43/10/0; further support came from ASI-INAF contracts
    I/05/07/0 and I/023/09/0.  
  GDZ acknowledges support from SFTC
  (UK) via the Advanced Fellowship Programme.
    SOHO is a project of international cooperation between ESA and NASA.
    CDS was built and operated by a consortium led by the
    Rutherford Appleton Laboratory (RAL), which includes
    UCL/Mullard Space Science Laboratory, NASA/ Goddard
    Space Flight Center, Max Planck Institute for Extraterrestrial
    Physics, Garching, and Oslo University.
    Finally, we wish to thank the referee for useful comments and suggestions
    on the manuscript.
\end{acks}

\end{article} 
\end{document}